\documentclass{aa}
\usepackage[varg]{txfonts}
\usepackage{graphicx}
\usepackage[dvipsnames]{xcolor}
\usepackage{booktabs} 

\usepackage{hyperref}
\hypersetup{colorlinks,linkcolor={blue},citecolor={blue},urlcolor={red}}
\usepackage{changes}
\usepackage{siunitx}
\usepackage{natbib}

\usepackage{amssymb}
\usepackage{amsmath} 

\usepackage{commath}

\usepackage[utf8]{inputenc} 
\usepackage[T1]{fontenc} 

\newcommand{\change}[1]{{#1}}
\newcommand{\newchange}[1]{{#1}}

\begin{document}

\title{Linking planetesimal and dust content in protoplanetary disks via a local toy model}
\titlerunning{Linking planetesimal and dust content in protoplanetary disks}

\author{Konstantin Gerbig \inst{\ref{inst:mpia},\ref{inst:ucsc}}, Christian T. Lenz \inst{\ref{inst:mpia}}, and Hubert Klahr\inst{\ref{inst:mpia}}}
\institute{Max Planck Institute for Astronomy, Königstuhl 17, 69117 Heidelberg, Germany, \email{gerbig@mpia.de}\label{inst:mpia}\and Department of Astronomy and Astrophysics, University of California, Santa Cruz, CA 95064, USA \label{inst:ucsc}}
\date{Received 15/02/2019 Accepted 30/07/2019}


\abstract{If planetesimal formation is an efficient process, as suggested by several models involving gravitational collapse of pebble clouds, then, before long, \change{a significant part of the primordial dust mass} should be absorbed in many \change{km} sized objects. A good understanding of the total amount of solids in the disk around a young star is crucial for planet formation theory. But as the mass of particles \change{above} the mm size \change{cannot} be assessed observationally, one must ask how much mass is hidden in bigger objects.}
{We perform 0-d local simulations to study how the planetesimal to dust and pebble ratio is evolving in time and to develop an understanding of the potentially existing mass in planetesimals for a certain amount of dust and pebbles at a given disk age.}
{\change{We perform a parameter study based on a model considering} dust growth, planetesimal formation and collisional fragmentation of planetesimals, \change{while neglecting} radial transport processes.}
{While at early times, dust is the dominant solid particle species, there is a phase during which planetesimals make up a significant portion of the total mass starting at approximately $10^4 - 10^6$ yr. The time of this phase and the maximal total planetesimal mass strongly depend on the distance to the star $R$, \change{the initial disk mass}, and the efficiency \change{of planetesimal formation} $\epsilon$. \change{Planetesimal collisions are more significant in} more massive disks, \change{leading to lower relative planetesimal fractions compared to less massive disks.} \change{After approximately $10^6$ yr, our model predicts planetesimal collisions to dominate, \newchange{which} resuppl\newchange{ies} small particles.}}
{\newchange{In our model, planetesimals form fast and everywhere in the disk.} \change{For a given $\epsilon$,} we were able to relate the dust content and mass of a given disk to its planetesimal content, providing us with some helpful basic intuition about \change{mass distribution of solids} and its dependence on underlying physical processes.}


\maketitle


\section{Introduction}

Planetesimals are \change{the smallest} solid objects that are bound by their own gravitational attraction in lieu of material binding forces. Since planetesimals \change{are believed to} constitute the building blocks of planets \change{\citep[e.g.][]{Kokubo2012}}, studying the physical processes involved in their formation and evolution within a protoplanetary disk around a young star is of great importance for planet-formation theory and thus also for the question of the origin of our Solar System's planets. Understanding the solid material available in a protoplanetary disk is a key aspect of this issue. However, observations are limited in what particle sizes can be detected. While for instance the Atacama Large Millimeter/submillimeter Array allows to determine the amount of mm-\change{ and  \si{\micro}m-sized particles via observations in the} scattered light \change{\citep[see e.g.][]{Andrews2009, Andrews2013, Ansdell2016}}, it is unknown how much solid material is in the size of planetesimals since they do not show characteristic absorption features and are too small to be be detected individually.
Models of planetesimal formation \change{\citep[e.g.][]{Drazkowska2016, Drazkowska2017} } based on the dust evolution model presented by \citet{Birnstiel2012} show rapid transformation of solid material from dust via pebbles into planetesimals. The ideal scenario of a model combining dust growth, pebble formation and planetesimal formation with evolution models \change{such as}  \citet{Morbidelli2009} or \citet{Kobayashi2016} would be numerically very expensive, because size ranges would be significantly larger than what is currently covered \change{in evolution models that include planetesimal interactions \citep[e.g.][]{Kokubo2002, Leinhardt2009, Levison2012, SanSebastian2019}}. Therefore, before starting this endeavor, we perform 0-d local simulations to study how the planetesimal to dust and pebble ratio is evolving in time. Hereby, we consider dust evolution based on the two-population model by \citet{Birnstiel2012}, pebble flux-regulated planetesimal formation after \citet{Lenz2019}, and collisional fragmentation of planetesimals to resupply small particles. Our aim is to develop an understanding of the potentially existing mass in planetesimals for a certain amount of dust and pebbles, by exploring \change{two} scenarios and \change{various} parameters.

In our model we ignore spatial transport of material\change{, which} is only important if inward drift of pebbles occurs on shorter timescales than their transformation into planetesimals. If the latter process is sufficiently efficient, spatial transport becomes unimportant. Further, we assume a constant gas column density by ignoring \change{viscous gas evolution \citep{Luest1952, Pringle1981,  Hueso2005, Birnstiel2010} or photoevaporation \citep[see e.g.][]{Ercolano2009, Owen2011, Nakatani2018, Picogna2019}}.

This would be of particular importance during the late stages of the disk evolution and in the inner disk regions. \newchange{We therefore limit the run time of our simulations to $10^7$~yr to be in line with typical lifetimes of protoplanetary disks \citep{Hernandez2007, Mamajek2009, Fedele2010, Pfalzner2014}.}

\change{The model principles are} described in detail in Sect. \ref{section:principles}, in particular the three processes \change{of} dust growth, planetesimal formation and planetesimal collisions.  \change{They} lead to a set of differential equations\change{, which we present for two scenarios in Sect.~\ref{section:numerical_setup}.} \change{In addition, we there discuss our numerical setup, i.e. our parameters, initial conditions and simulation run time}. We present our results obtained from a parameter study in Sect. \ref{section:results}\change{, where} we further estimate how the total mass in the disk is distributed among the three species dust, pebbles, and planetesimals. \change{In Sect. \ref{section:limitations}, we discuss model limitations.} Finally, we summarize and \change{conclude} in Sect. \ref{section:summary}.


\section{Model principles}
\label{section:principles}

\subsection{Protoplanetary disk setup}
\label{section:ppdisk_setup}

A protoplanetary disk \change{initially} contains gas and dust circulating around a central star with mass $M_\mathrm{star}$. We use the concept of column densities, which is commonly used in astrophysics when describing accretion disks. In cylindrical coordinates ($R, \phi, z$) it is defined as the integral of the volumetric mass density $\rho$ \change{of gas or solids} integrated along the vertical path $z$ that goes through the disk
\begin{equation}
\Sigma(R,\phi) = \int_{-\infty}^{\infty}\rho (R, \phi, z)\mathrm{d}z.
\end{equation}
By assuming a cylindrical symmetric disk, one can eliminate the $\phi$-dependence of the column density.
If spatial transport of material is neglected, conservation of mass implies that also the total column density \change{of all solid material $\Sigma_\mathrm{solids}$} must be conserved
\begin{equation}
\frac{\partial\Sigma_\mathrm{\change{solids}}(R,t)}{\partial t} = 0.
\end{equation} \change{Initially, the total column density of solids is made up purely by dust. We relate the dust column density $\Sigma_\mathrm{dst}$ to the gas column density $\Sigma_\mathrm{g}$ with the dust-to-gas ratio $\epsilon_\mathrm{dg}$, which we assume to be $R$-independent. The total column density of all material $\Sigma_\mathrm{total}$, including solids and gas, is then given by
\begin{equation}
\label{eq:Sigma_total}
\Sigma_\mathrm{total} = \Sigma_\mathrm{g} + \Sigma_\mathrm{solids} = \Sigma_\mathrm{g} + \Sigma^0_\mathrm{dst} = \left(1+ \epsilon^0_\mathrm{dg}\right)\Sigma_\mathrm{g},
\end{equation}
where we introduced the initial dust-to-gas ratio
\begin{equation}
\epsilon^0_\mathrm{dg} = \frac{\Sigma^0_\mathrm{dst}}{\Sigma_\mathrm{g}} = 10^{-2},
\end{equation}
as found by \citet{Savage1972} or \citet{Draine2007} in the interstellar medium (ISM). The gas profile is non-evolving in our model, hence also the total column density $\Sigma_\mathrm{total}$ in Eq. \eqref{eq:Sigma_total} must be constant in time. The initial column density} depends on $R$ according to a self-similar initial profile derived by \citet{Lynden1974},
\begin{equation}
\label{eq:initial_dust_profile}
\Sigma_\mathrm{total}(R) = C_{\Sigma} \left(\frac{R}{R_\mathrm{C}}\right)^{\change{-}n} \exp\left[-\left(\frac{R}{R_\mathrm{C}}\right)^{2-n}\right],
\end{equation}
where $R_\mathrm{C}$ is the so-called characteristic radius, marking \change{the transition of the power law to the exponential}. $C_{\Sigma}$ is a normalization constant chosen such that the disk contains a total mass of solids and gas \change{of} $M_\mathrm{disk}$. For our simulations we choose $R_\mathrm{C} = 40 \mathrm{\ AU}$, roughly corresponding to Kuiper Belt's location in the Solar System. 

\change{For a vertically isothermal disk with temperature $T$, the sound speed $c_\mathrm{s}$ of molecular hydrogen is given by
\begin{equation}
c_\mathrm{s} = \sqrt{\frac{k_\mathrm{B} T}{2 m_\mathrm{p}}},
\end{equation}
with Boltzmann constant $k_\mathrm{B} = 1.3807 \cdot 10^{-16} \mathrm{erg}\,\mathrm{K}^{-1}$, and proton mass $m_\mathrm{p} = 1.673\cdot10^{-24} \mathrm{\ g}$. Further, the vertical pressure scale height of the gas disk $h_\mathrm{g}$ is defined as
\begin{equation}
    h_\mathrm{g} = \frac{c_\mathrm{s}}{\Omega},
\end{equation} 
which originates from considering vertical hydrostatic balance \citep{Weizsacker1948}. Here, we introduced the Keplerian angular velocity $\Omega$, which is derived from balance of centrifugal and gravitational acceleration
\begin{equation}
\Omega = \sqrt{\frac{G M_\mathrm{star}}{R^3}},
\end{equation}
where $G \approx  6.674 \cdot 10^{-8}\ \mathrm{cm}^3\,\mathrm{g}^{-1}\,\mathrm{s}^{-2}$. We set the
exponent of the polynomial decline in Eq, \eqref{eq:initial_dust_profile} to $n = 1$ after the radial viscosity profile of turbulent disks presented by \citet{Shakura}:
\begin{equation}
    \nu = \alpha c_\mathrm{s}h_\mathrm{g}\propto \frac{c_\mathrm{s}^2}{\Omega} \propto T\cdot R^{3/2}\propto R^{3/2-q}=R^1,
\end{equation}
\newchange{to which the column density is inversely proportional in the inner disk. We assume a }temperature profile  \newchange{of} \citep{Chiang1997}
\begin{align}
T = T_\mathrm{star}\left(\frac{R_\mathrm{star}}{R}\right)^{q}\alpha_\mathrm{irr}^{0.25},
\end{align}
where we choose stellar properties of $R_\mathrm{star} = 1.25 \cdot R_{\odot} \approx 1.25 \cdot 7 \cdot 10^{10} \mathrm{\ cm}$ and $T_\mathrm{star} =  4000 \mathrm{\ K}$ \citep{Beckwith1990}, $q = 0.5$ after \citet{Chiang1997} and an irradiation angle of $\alpha_{\mathrm{irr}}= 0.1$ \citep[e.g.][]{Pfeil2019}. This profile is valid for disks where heating is dominated by radiation from the star, in lieu of viscous heating in accretion disks \citep{Pringle1981}, which we assume to be insignificant in comparison.}

\change{
Lastly, the dimensionless parameter $\alpha$ quantifies disk turbulence. Typical values are in the range $10^{-4} \leq \alpha \leq 10^{-2}$ as concluded from simulations \citep{Johansen2005, Dzyurkevich2010, Nelson2013, Flock2017} or observations such as \citet{Andrews2009, Flaherty2017}.}


\subsection{Dust growth}
\label{section:dust_growth}

\citet{Birnstiel2012} showed that growth of dust particles in protoplanetary disks can be sufficiently described by representing the entire dust population by \change{only} two sizes. The small size (henceforth referenced to as \textit{dust}) represents those particles which are tightly coupled to the gas \change{via aerodynamic friction}. The \change{larger} size (henceforth referenced to as \textit{pebbles}) represents particles \change{subject to \newchange{a substantial} radial drift}. These species are characterized by their respective Stokes numbers, a crucial quantity for describing the \change{aero}dynamic behavior of particles. \change{For $\mathrm{St}\gg 1$, they are decoupled from the gas, while for $\mathrm{St}\ll 1$, they are strongly affected by gas dynamics}. Assuming monodisperse coagulation \citep{Stepinski1997, Kornet2001, Brauer2008, Birnstiel2012} the size $a$ of a dust particle at a certain time $t$ can be described by an expression of exponential growth
\begin{equation}
\label{eq:dustgrowthsizes}
a = a_0 \exp\left(\frac{t}{\tau_\mathrm{growth}}\right),
\end{equation}
where $a_0$ is an initial reference size and $\tau_\mathrm{growth}$ a certain growth timescale. Equation \eqref{eq:dustgrowthsizes} is only valid for constant $\epsilon_{\mathrm{dg}}$, which we assume to be approximately true, as the resulting error is insignificant in comparison to other assumptions in model. We simplify the two-population model by assigning the dust species to $a_0 =$ \SI{0.1}{\micro\meter} after \citet{Mathis1977} and the pebble species to $a = a(\mathrm{St}_\mathrm{pbb})$. 

\change{As the size of dust grains in our model remains smaller than the mean free path of the gas}, the Epstein regime \newchange{of gas drag} is valid and the molecular nature of the gas has to be considered \citep{Epstein1924}. \change{Near the midplane we can then follow \citet{Birnstiel2012} and calculate $\mathrm{St}_\mathrm{dst}$ via 
\begin{equation}
\label{eq:St_epstein}
\mathrm{St}_\mathrm{dst} = \frac{\pi}{2}\frac{a_0 \rho}{\Sigma_\mathrm{g}},
\end{equation} where $\rho$ is the mass density of dust. We choose $\rho = 1.2 $ g cm$^{-3}$, which was found by \citet{Carry2012} for asteroids.}

\change{Fragmentation \citep{ Blum1993, Blum2008} and radial drift \citep{Klahr2006, Birnstiel2012} may limit the maximum pebble size. In the inner disk, the former is typically more significant, whereas in the outer disk, the latter typically is more important \citep{Birnstiel2012, Lenz2019}. As these limits and their evolution are not the focus of this paper, we will simplify and treat} the Stokes number of pebbles as a fixed parameter of our model with $\mathrm{St}_\mathrm{pbb} = 0.1$ \change{approximately corresponding to the maximum Stokes number of the larger particle species as found by \citet{Birnstiel2012}. Implications are discussed in \newchange{Appendix}~\ref{sec:pebble_stokes_number}.} 

\change{Following \citet{Birnstiel2012}}, we \change{estimate} the dust-to-pebbles growth timescale \change{in the midplane} with
\begin{equation}
\label{eq:adjustedgrowthtimescale}
\tau_\mathrm{growth} \simeq \ln{\left(\frac{\mathrm{St}_\mathrm{pbb}}{\mathrm{St}_\mathrm{dst}}\right)} \frac{1}{\epsilon_\mathrm{dg} \Omega}.
\end{equation}
\change{The approximation in Eq. \eqref{eq:adjustedgrowthtimescale} holds if relative dust velocities are set by disk turbulence $\Delta v_\mathrm{dst} \approx c_s \sqrt{3 \alpha \mathrm{St}_\mathrm{dst}}$ with $\mathrm{St}_\mathrm{dst} \ll 1$, i.e. for sufficiently large $\alpha$-values.}

\change{The dust to pebble growth rate is calculated via
\begin{equation}
\label{eq:dust_to_pebble_growth_rate}
\dot{\Sigma}_\mathrm{growth} = \frac{\Sigma_\mathrm{dst}}{\tau_\mathrm{growth}} = \ln{\left(\frac{\mathrm{St}_\mathrm{dst}}{\mathrm{St}_\mathrm{pbb}}\right)} \epsilon_\mathrm{dg} \Omega\Sigma_\mathrm{dst} \propto \Sigma_\mathrm{dst}^2.
\end{equation}
}
\change{The differential equation \begin{equation}
    \dot{\Sigma}_\mathrm{dst} = - \dot{\Sigma}_\mathrm{growth}
\end{equation} illustrates the decrease of the local dust supply due to particle growth to pebble size. It is solved analytically by
\begin{equation}
\label{eq:analytical_solution}
\Sigma_\mathrm{dst}(t) = \left[C\cdot t+\frac{1}{\Sigma_\mathrm{dst}^0}\right]^{-1}.
\end{equation}
Here, $\Sigma_\mathrm{dst}^0$ is the initial dust column density from the profile in Eq. \eqref{eq:initial_dust_profile} and
\begin{equation}
C := \frac{\Omega}{\Sigma_\mathrm{g}\ln{\left(\frac{\mathrm{St}_\mathrm{pbb}}{\mathrm{St}_\mathrm{dst}}\right)}}
\end{equation}
is constant in time for a fixed $\mathrm{St}_\mathrm{pbb}$.}


\subsection{\change{Pebble flux-regulated} planetesimal formation}
\label{section:pls_formation}

A gas parcel at a certain distance from the central star is in a force balance between gravitational, centrifugal and thermal pressure forces. A solid particle does not feel this outward oriented pressure force and should therefore move on a Keplerian orbit, whereas the gas moves on a sub-Keplerian orbit. Because of \change{aerodynamic friction}, solid particles lose angular momentum to the gas and spiral inward towards the central star. \change{After \citet{Nakagawa1986}, the radial drift velocity of pebbles can be derived by considering the equations of motions of particles and gas
\begin{equation}
\label{eq:vdrift}
v_\mathrm{drift} = \frac{\mathrm{St}_\mathrm{pbb}}{\mathrm{St}_\mathrm{pbb}^{2} + \left(1 + \epsilon_\mathrm{dg}\right)^2} \frac{h_\mathrm{g}}{R} \gamma c_\mathrm{s},
\end{equation}
where we introduced the exponent of the gas pressure power law
\begin{align}
\gamma = \frac{\partial \ln P}{\partial \ln R},
\end{align}
which is for $R\ll R_\mathrm{C}$ and $T\propto R^{-1/2}$ given by $\gamma = -2.75$. Due to the assumption of a constant pebble Stokes number, the drift velocity of pebbles in Eq. \eqref{eq:vdrift} remains almost constant as only $\epsilon_{\mathrm{dg}}$ changes in time.}

One finds that pebbles with $\mathrm{St}_\mathrm{pbb} \approx 10^{-1}$ quickly drift inward towards the central star on very short timescales. Such a particle at $10$ AU can be expected to reach the star in under $10^4$ yr. If pebbles can not grow an order of magnitude in a shorter time than this, they fall victim to evaporation in the inner disk. Particle growth also leads to higher relative velocities, so that upon collision the bodies tend to break up rather than stick together \citep{Blum1993, Chokshi1993, Blum2008}. Therefore, further coagulation from pebbles to planetesimals seems unlikely \change{\citep{Homma2018}} , though not impossible as for example \change{shown} in \citet{Kataoka2013}.

Our work focuses on a different way to overcome the fragmentation and drift barrier, as first suggested by \citet{Safronov1969} \change{and later \citet{Goldreich1973}}. Here, when particles \newchange{settle} towards the midplane, they undergo a gravitational instability and form planetesimals in a spontaneous event. \citet{BalbusHawley1998} found that magneto-rotational instability in the disk forms short-lived turbulent eddies, vortices and pressure bumps which \newchange{may} trap pebble-sized particles \citep{Johansen2005}. Further, streaming instability \newchange{\citep{Youdin2005, Squire2018, Umurhan2019}}, where the feedback of particles onto gas is crucial, may be able to create \newchange{significant} particle over-densities \newchange{\citep{Johansen2007ApJ, Carrera2015, Simon2016, Nesvorny2019}}. \change{If} \newchange{a} particle cloud \change{is not ripped apart by the star's tidal forces, i.e. has reached Hill-density,} it can collapse and form a planetesimal as discussed in \citet{Johansen2006} and \citet{Johansen2007}. \change{The initial size of the resulting planetesimal \newchange{may} then \newchange{be} given by the balance of contraction and particle diffusion timescale \citep{KlahrSchreiber2015}.}
Instead of considering different types of traps and examining physical properties of the disk in detail, we will \change{follow \citet{Lenz2019} and} describe the trapping mechanism with \change{the} help of parameters, most important of which is the planetesimal formation efficiency $\epsilon$. We will also refer to $\epsilon$ as the trap (or trapping) efficiency, but it is important to keep in mind, that it quantifies not only how efficient the trap can accumulate pebbles, but also how efficient \change{trapped} pebbles can be \change{converted} into planetesimals. The trap distance $d$ quantifies \newchange{typical} radial \newchange{separation} of the trap and $\tau_\mathrm{trap}$ a trap's typical lifetime. \change{We set $d = 5h_\mathrm{g}$ and $\tau_\mathrm{trap}$ to 100 orbits, as found numerically by \citet{Dittrich2013, Manger2018}}.

Following \citet{Lenz2019}, \change{pebbles are transformed into planetesimals over the conversion length
\begin{equation}
l := \frac{d}{\epsilon}.
\end{equation}} We \change{further} express the planetesimal formation rate by
\begin{equation}
\label{eq:formationrate}
\dot{\Sigma}_\mathrm{form} = 
\frac{ |v_\mathrm{drift}|}{\change{l}}\Sigma_\mathrm{pbb}.
\end{equation}
\change{Here, we ignore the contribution of any dust that might also be trapped in the collapsing cloud. Dust is slowed down more significantly than pebbles, as the sedimentation (or contraction) timescale $\tau_\mathrm{sed}$ for $\mathrm{St} < 1$ particles increases due to aerodynamic friction on the order of
\begin{equation}
    \tau_\mathrm{sed} \approx  \frac{\tau_\mathrm{ff}}{\mathrm{St}},
\end{equation}
where $\tau_\mathrm{ff}$ is the free-fall timescale \citep{Shariff2015, KlahrSchreiber2015}. Hence, we expect pebbles to dominate the mass contribution of the collapsing cloud. 
}

Of course, \change{the recipe in Eq. \eqref{eq:formationrate}} is only valid once the flux of pebbles through the trap has reached a critical value, in other words once the trap has accumulated enough mass to form at least a single planetesimal with mass $m_\mathrm{pls}$ \change{\citep{Lenz2019}}:
\begin{equation}
M_\mathrm{trapped}=2\pi R \epsilon\tau_\mathrm{trap} \int_{\mathrm{St}_\mathrm{min}}^{\mathrm{St}_\mathrm{max}} |v_\mathrm{drift}|\Sigma_\mathrm{\change{pbb, St}}\mathrm{dSt} \geq m_\mathrm{pls},
\end{equation}
\change{where $\Sigma_\mathrm{pbb, St}$ is the pebble column density per unit Stokes number.} Since in this model, we assume all pebbles to have the same constant Stokes number, the integral vanishes and the condition simplifies to
\begin{equation}
\label{eq:M_trapped}
M_\mathrm{trapped}=2\pi R \epsilon\tau_\mathrm{trap} |v_\mathrm{drift}|\Sigma_\mathrm{pbb} \geq m_\mathrm{pls}.
\end{equation}
\change{If Eq. \eqref{eq:M_trapped} is not fulfilled, $\dot{\Sigma}_\mathrm{form}$ is set to 0.}
\change{We deduce the fixed parameter $m_\mathrm{pls}$ from the initial planetesimal size, which can be estimated by studying Asteroid Belt and Kuiper Belt objects ---} relics of the planet formation process in the Solar System. Observations of the cumulative \change{size} distribution of bodies in these regions show a \change{decrease} of slope \change{for increasing sizes} at roughly $100$ km \citep{Bottke2005, Nesvorny2011, Fraser2014, Delbo2017}. It was concluded that this kink cannot be obtained by collisional evolution alone and instead originates from the primordial size distribution. \citet{Morbidelli2009} proposed that planetesimals were born with a minimum diameter of $80$ km, which \change{is} proven to be possible \citep{Johansen2007, Cuzzi2008, KlahrSchreiber2015}. Assuming homogeneous and spherical planetesimals with mass density $\rho = 1.2 \mathrm{\ g}\,\mathrm{cm}^{-3}$, the planetesimal mass is approximated with
\begin{equation}
m_\mathrm{pls} = \frac{4\pi}{3} \rho r_\mathrm{pls}^3,
\end{equation}
where $r_\mathrm{pls} = 50$ km is the planetesimal radius.  \change{We note that this recipe for planetesimal formation also works for smaller masses, as the pebble flux is typically super-critical \citep{Lenz2019}.}

Our model does not include a mass grid and therefore has to also exclude growth of planetesimals to larger sizes, meaning the planetesimal population will exclusively contain 100 km sized objects.


\subsection{Planetesimal collision \change{model}}
\label{section:pls_collisions}

Collisions between two planetesimals are complex physical processes with many potential outcomes. The simplified picture used in this paper's model imagines the column density of the two colliding planetesimals to be redistributed among fragments of different sizes upon a collisional event. \change{Collisions are \newchange{thus} assumed to be} \newchange{destructive, as the total planetesimal column density in the system decreases upon planetesimal collision.} The most convenient approach is to apply an inverse power law for the function of the specific number column density $n_m(m)$, which describes the fragment distribution by assuming a collisional cascade, as seen in e.g. \citet{Zvyagina1973}. The specific number column density reads
\begin{equation}
\label{eq:col_powerlaw}
n_m(m) = C_n \cdot m^{-\xi},
\end{equation}
where $C_n$ is \change{a} normalization constant and $\xi = 1.83$  as found by \citet{Dohnanyi1969} \change{by analytically investigating the collective dynamical interactions of asteroids for inelastic collisions and fragmentation.} \change{This value was also found experimentally via collision experiments with basalt rocks \citep{Fujiwara1977}. The slope of the power law incorporates specifics of a collisional event, such as impact velocities, which} \newchange{\citet{Dohnanyi1969} assumes to be} \change{on the order of kilometers per second.} \newchange{We point out, that this assumption is overestimating typical planetesimal impact velocities in protoplanetary disks by about 2 orders of magnitude \citep{Wetherhill_1993, Morbidelli2009}. Although, $\xi = 1.83$ is robust against variations of physical parameters like the impact velocity \citep{Dohnanyi1969}, and our collision model is robust against variations of $\xi$, we note that we therefore also overestimate the steepness of the power law.} \change{We derive the normalization $C_n$ in Appendix~\ref{sec:collision_normalize}.}

\change{As we are only interested in an average size distribution of fragments, we express} the column density that is redistributed into objects with masses between \change{$m_i$ and $m_j$ via the integral $\int_{m_i}^{m_j}n_m m dm$.} 

\change{Now,} the column density redistribution ratio upon a collisional event \change{can be estimated} by defining three mass intervals corresponding to the three species dust, pebbles and planetesimals, \change{i.e. for dust $m_0 \leq m < m_1$, for pebbles $m_1 \leq m < m_2$ and for planetesimals $m_2 \leq m \leq M$.} \change{The derivations of $m_1$ and $m_2$ are shown in Appendix \ref{section:transition_masses}.}

With these requisites, one can solve the integral $\int_{m_i}^{m_j}n_m m dm$ for each population, which leads to three fractions $p_\mathrm{dst}, p_\mathrm{pbb}$ and $p_\mathrm{pls}$  corresponding to the column density participating in a single collision that \change{is} redistributed to dust, pebbles and planetesimals, respectively. \change{For typical disk parameters (see tab. \ref{tab:default_set}), $m_1, m_2 \ll m_\mathrm{pls}$. With $\xi = 1.83$, the mass in fragments is therefore significantly dominated by objects which we still consider to be planetesimals.}

The \change{effect} of planetesimal collisions on the surface density evolution is affected not only by the outcome of a single collision, but also on the frequency of planetesimal encounters in a protoplanetary disk, quantified by planetesimal collision timescale and rate. \change{We} consider the mean free path $\lambda_\mathrm{mfp}$ between collisions. If a single particle $i$ is moving through a cluster of other particles $j$ with number density $n_j$ and cross section for collisions between particles $i$ and $j$ is $\sigma_{ij}$, then \citep{Birnstiel2016}
\begin{equation}
\lambda_\mathrm{mfp} = \frac{1}{n_j \sigma_{ij}}.
\end{equation}
The collision timescale $\tau_{\mathrm{col},i}$ is defined as the average time \change{during which} particle $i$ experiences one encounter
\begin{equation}
\label{eq:tau_col_general}
\tau_{\mathrm{col},i} = \frac{\lambda_\mathrm{mfp}}{\Delta  v_{ij}} = \frac{1}{n_j \sigma_{ij} \Delta  v_{ij}},
\end{equation}
where $\Delta  v_{ij}$ is the relative velocity of the particles $i$ and $j$ to each other. Since all planetesimals in our model are equal-sized, the cross section two planetesimals $i$ and $j$ with radius $r_\mathrm{pls}$, is given by the gravitational cross section $\sigma_\mathrm{grav}$ \citep{Safronov1969}, which considers both the geometry of the system but also gravitational focusing:
\begin{equation}
\label{eq:crosssection_grav}
\sigma_\mathrm{grav} = 4\pi r_\mathrm{pls}^2 \left[1 + \left(\frac{v_\mathrm{esc}}{\Delta  v_{ij}}\right)^2\right],
\end{equation}
where the escape velocity $v_\mathrm{esc}$ of the planetesimals is
\begin{equation}
    v_\mathrm{esc} = \sqrt{\frac{2 G m_\mathrm{pls}}{r_\mathrm{pls}}}.
\end{equation} By approximating the vertical planetesimal distribution with a Gaussian with \change{a root mean square width} of $h_\mathrm{pls} = v_\mathrm{hill}\Omega^{-1}$ \citep{Goldreich2004a}, the number density of planetesimals can be written as
\begin{equation}
n_\mathrm{pls} = \frac{\Sigma_\mathrm{pls} \Omega}{\sqrt{2\pi}m_\mathrm{pls}v_\mathrm{hill}}.
\end{equation}
Here, we introduced the Hill velocity $v_\mathrm{hill}$, which is the velocity of particle $i$ relative to particle $j$ when they are just close enough to each other that the total gravitational force acting on them is not dominated by the mass of the central star $M_\mathrm{star}$, but instead their own masses $m_i$ and $m_j$, i.e. when entering their respective Hill spheres. For equally massive planetesimals the hill velocity reads \change{\citep{Hill1878}}
\begin{equation}
v_\mathrm{hill} = R \Omega\left(\frac{2m_\mathrm{pls}}{3M_\mathrm{star}}\right)^{1/3}.
\end{equation}
Plugging the above expressions into \eqref{eq:tau_col_general} we get for the collision timescale of equal-sized planetesimals
\begin{equation}
\label{eq:collisiontimescale}
\tau_\mathrm{col} = \frac{\sqrt{2\pi}m_\mathrm{pls}v_\mathrm{hill}}{\sigma_\mathrm{grav}\Sigma_\mathrm{pls}\Omega\Delta  v_{ij}}.
\end{equation}
We follow \citet{Morbidelli2009} and insert the Hill velocity for the relative velocity of planetesimals $\Delta  v_{ij}$. We note that the relative velocity appears once directly in \eqref{eq:tau_col_general} and again as its inverse squared in \eqref{eq:crosssection_grav}, leading to the perhaps counter intuitive implication of higher relative velocities causing longer collision timescales.
The rate of planetesimal collisions $\dot{\Sigma}_\mathrm{col}$ must also depend on the available planetesimal surface density. It is given by
\begin{equation}
\dot{\Sigma}_\mathrm{col} = \frac{\Sigma_\mathrm{pls}}{\tau_\mathrm{col}} \propto \Sigma_\mathrm{pls}^2 .
\label{eq:collisionrate}
\end{equation}


\section{\change{Numerical model setup}}
\label{section:numerical_setup}

\begin{table*}[t]
	\centering
	\caption[]{Default parameter set }
	\label{tab:default_set}
	\begin{tabular*}{\linewidth}{lllll}
		\toprule
		Parameter & Value & Unit  & Parameter name & \change{References}\\
		\midrule
		$R$ & $10$ & AU  & Distance to star & Arbitrarily set to Saturn's position in Solar System\\
		$M_\mathrm{disk}$ & $0.02$ & $ M_{\mathrm{star}}$ &  Disk mass (gas + solids)& MMSN \citep{Weidenschilling1977MMSN}\\
		$\epsilon$ & $0.01$ &  & Trap efficiency  & Chosen arbitrarily\\
		$\mathrm{St}_\mathrm{pbb}$ & $0.1$ &  & Pebble Stokes number& Maximum Stokes number in \citet{Birnstiel2012}\\
		\change{$M_\mathrm{star}$} & $1$& $M_{\odot}$& Stellar mass & Set to mass of Sun \\
		\change{$T_\mathrm{star}$} & $4000$ & K & Stellar effective temperature & \citet{Beckwith1990, Chiang1997}\\
		\change{$R_\mathrm{star}$ }& $1.25$& $R_{\odot}$ & Stellar radius & Fiducial value for pre-main sequence stars \\
		\change{$\alpha_\mathrm{irr}$} & $0.1$ & & Irradiation angle & E.g. \citet{Pfeil2019} \\
		$\epsilon_\mathrm{dg}^0$ & $0.01$ &  &  Initial dust-to-gas ratio & \citet{Savage1972, Draine2007}\\
        $R_\mathrm{C}$ & $40$ & AU &  Characteristic radius of initial dust profile & Set to Kuiper Belt's location in Solar System\\
		$n$ & $1$ &   & Power-law exponent of initial dust profile & After radial viscosity profile in \citet{Shakura}\\
		$a_0$ & $10^{-4}$ & cm &  Dust grain radius & \citet{Mathis1977}\\
		$r_\mathrm{pls}$ & $50$ & km & Planetesimal radius & \citet{Morbidelli2009, KlahrSchreiber2015} \\
		$\rho$ & $1.2$ & $\mathrm{g}\,\mathrm{cm}^{-3}$ &  Volume material density & \citet{Carry2012}\\
		$d$ & $5 $ & $h_\mathrm{g}$  & \change{Trap distance} &\citet{Dittrich2013}\\
		$\tau_\mathrm{\newchange{trap}}$ & $100$   & Orbits & Trap lifetime & \citet{Dittrich2013, Manger2018} \\
		\bottomrule
	\end{tabular*}
\end{table*}

\subsection{Balance equations \change{for two scenarios}}
\label{section:balance_eqs}

The \change{evolution of the column densities due to the dust growth to pebbles, planetesimal formation and planetesimal collisions} can be assembled to \change{two} systems of coupled differential equations \change{corresponding to two different scenarios}. \change{The rates for the model processes are represented by} \change{sink} and source terms:
\begin{subequations}
\begin{align}
\label{eq:scenario1diffeq1}
\dot{\Sigma}_\mathrm{pls} &= \dot{\Sigma}_\mathrm{form}  - \left(1 - p_\mathrm{pls}\right) \dot{\Sigma}_\mathrm{col},\\
\label{eq:scenario1diffeq2}
\dot{\Sigma}_\mathrm{pbb} &= \dot{\Sigma}_\mathrm{growth} - \dot{\Sigma}_\mathrm{form}  + p_\mathrm{pbb} \dot{\Sigma}_\mathrm{col},\\
\label{eq:scenario1diffeq3}
\dot{\Sigma}_\mathrm{dst} &= - \dot{\Sigma}_\mathrm{growth} + p_\mathrm{dst} \dot{\Sigma}_\mathrm{col}.
\end{align}
\end{subequations}
Since $\dot{\Sigma}_\mathrm{growth}$ and $\dot{\Sigma}_\mathrm{form}$ each appear once as a \change{sink} and a source term and $p_\mathrm{dst} + p_\mathrm{pbb} + p_\mathrm{pls} = 1$ per definition, the total column density in the system is conserved in time
\begin{equation}
    \frac{\partial}{\partial t} \left(\Sigma_\mathrm{dst} + \Sigma_\mathrm{pbb} + \Sigma_\mathrm{pls}\right) = 0.
\end{equation}
In this configuration, we expect a column density steady state to occur. 

\newchange{Another} view arises when one considers that planetesimal collisions \newchange{may} produce very compact fragments\newchange{, in lieu of fluffy aggregates}. In particular, \newchange{if planetesimal formation compactifies the material involved in the gravitational collapse}, then collisional dust \newchange{would be} much more dense than primordial \change{dust aggregates}. \change{Material properties of dust grains are important for the microphysics of growth \citep{Ormel2007, Paszun2009}}. For very compact dust particles, growth via sticking \newchange{may be more difficult than for fluffy aggregates, because} \change{they may not be able to absorb all of the collisional energy and consequently break apart or restructure rather than grow. This is especially the case} if they are not encased by an ice mantle \change{which could be destroyed in collisions, provided there is no recondensation.} \change{Further, even if two compact grains collide gently enough, they may be very loosely packed such that future collisions are more likely to be destructive.} 

\begin{figure}[t]
\centering
\includegraphics[width=\linewidth]{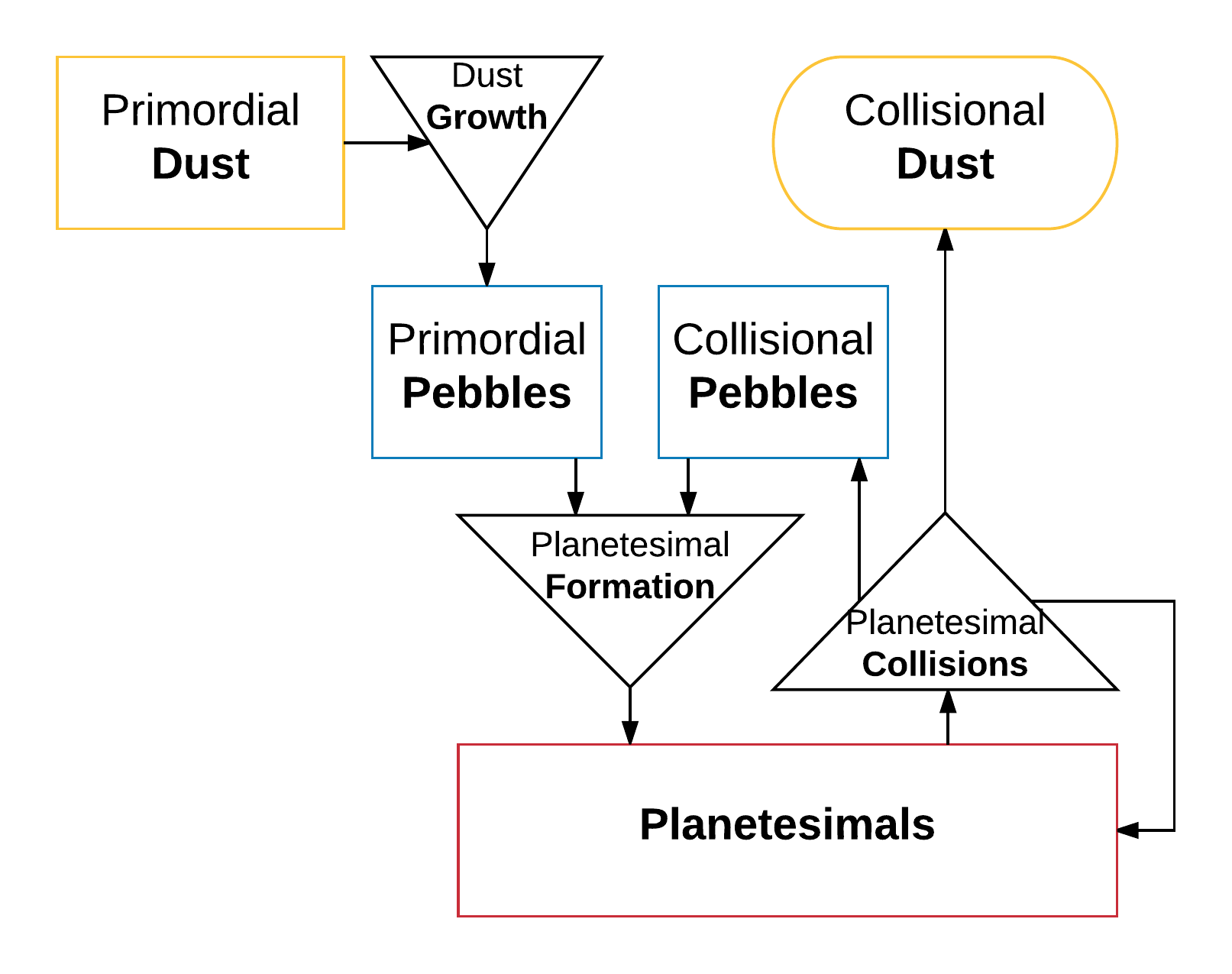}
\caption[]{Flowchart visualizing the different column density transfer processes in a scenario 2 configuration, where collisional dust can not grow. Scenario 1 would simply add an arrow connecting collisional dust to the dust growth triangle, \change{rendering the primordial populations equivalent to their collisional counterpart}.}
\label{fig:flowchart}
\end{figure}

\change{Therefore, we} differentiate dust and pebbles further into the subspecies \change{\textit{primordial}} and \change{\textit{collisional}} dust and pebbles respectively. \change{We note that primordial pebbles do not exist at the beginning of the simulation. Rather, they grow directly from primordial dust grains. This is in contrast to collisional pebbles, which originate from material that was previously comprised in planetesimals.} \change{We} introduce two additional differential equations
\begin{subequations}
\begin{align}
\label{eq:firstrateeq}
\dot{\Sigma}_\mathrm{pls} &= \dot{\Sigma}_\mathrm{form} - \left(1 - p_\mathrm{pls}\right)\dot{\Sigma}_\mathrm{col},\\
\dot{\Sigma}_\mathrm{pbb, prim} &= \dot{\Sigma}_\mathrm{growth} - f_\mathrm{pbb,prim}\dot{\Sigma}_\mathrm{form},\\
\label{eq:primdst}
\dot{\Sigma}_\mathrm{dst, prim} &= -  \dot{\Sigma}_\mathrm{growth},\\
\label{eq:colpbbdot}
\dot{\Sigma}_\mathrm{pbb, col} &= - f_\mathrm{pbb,col}\dot{\Sigma}_\mathrm{form} + p_\mathrm{pbb}\dot{\Sigma}_\mathrm{col},\\
\label{eq:coldst}
\dot{\Sigma}_\mathrm{dst, col} &= p_\mathrm{dst} \dot{\Sigma}_\mathrm{col},
\end{align}
where 
\begin{align}
    f_\mathrm{pbb,prim} &= \frac{\Sigma_{\mathrm{pbb, prim}}}{\Sigma_{\mathrm{pbb, prim}} + \Sigma_{\mathrm{pbb, col}}}\quad\text{and}\\
    f_\mathrm{pbb,col} &= 1 - f_\mathrm{pbb,prim}
\end{align}
\end{subequations}
correspond to the fraction of pebbles that is considered primordial and collisional respectively. This configuration is visualized per flowchart in Fig. \ref{fig:flowchart}. In contrast to \eqref{eq:scenario1diffeq1} - \eqref{eq:scenario1diffeq3}, collisional dust is too compact and dense to grow further, whereas collisional pebbles may participate in planetesimal formation again.

We will denote the configuration depicted in equations \eqref{eq:scenario1diffeq1}, \eqref{eq:scenario1diffeq2} and \eqref{eq:scenario1diffeq3}, where no differentiation between primordial and collisional dust and pebbles is performed, as scenario 1. Scenario 2 will describe the configuration in Fig. \ref{fig:flowchart}. Contrary to scenario 1, scenario 2 will not reach an equilibrium state as \eqref{eq:primdst} only contains a loss term, while Eq. \eqref{eq:coldst} only has a source term. \change{These two scenarios represent limiting cases. Intermediate scenarios where a fraction of collisional dust can grow, for example by coagulating with primordial dust, are also conceivable but not subject of this work.}

\subsection{\change{Initial conditions and default parameters}}
\label{section:inital_conditions}

Due to locality of the model, it is very easy to implement and computationally inexpensive. The code written specifically for this local model is based on the 2nd order Runge Kutta algorithm, a one-step procedure for solving differential equations with boundary conditions. \change{We set the initial primordial dust column density to
\begin{subequations}
\begin{align}
   \mathrm{Scenario \ 1: \ } & \Sigma_\mathrm{dst}^0 = \left(\frac{1}{\epsilon_\mathrm{dg}^0} + 1\right) \Sigma_\mathrm{total}(R),\\
\mathrm{Scenario \ 2: \ } & \Sigma_\mathrm{dst,prim}^0 = \left(\frac{1}{\epsilon_\mathrm{dg}^0} + 1\right) \Sigma_\mathrm{total}(R), 
\end{align}
\end{subequations}
according to the radial column density profile in Eq. \eqref{eq:initial_dust_profile}. All other populations are set to zero, i.e.
\begin{subequations}
\begin{align}
   \mathrm{Scenario \ 1: \ } & \Sigma_\mathrm{pbb}^0 = \Sigma_\mathrm{pls}^0 = 0,\\
\mathrm{Scenario \ 2: \ } & \Sigma_\mathrm{pbb,prim}^0 = \Sigma_\mathrm{pls}^0 = \Sigma_\mathrm{dst, col}^0 = \Sigma_\mathrm{pbb, col}^0 = 0.  
\end{align}
\end{subequations}
}

\change{Our simulation utilizes parameters introduced in Sect. \ref{section:principles} and summarized} in Table \ref{tab:default_set}. \change{In our parametric study, we vary $R$, $M_\mathrm{disk}$ and $\epsilon$.} The default value for the distance to the star $R = 10$ AU was chosen arbitrarily and corresponds roughly to the position of the planet Saturn in the solar system (semi-major axis of Saturn: 9.537 AU). $M_\mathrm{disk} = 0.02 \cdot M_{\odot}$ is roughly twice the mass of the Minimum Mass Solar Nebula (MMSN) as studied in \citet{Weidenschilling1977MMSN} \change{and \citet{Hayashi1981}}. The MMSN will be discussed in more detail later on, when looking at the influence of the disk mass on the simulation results. For the trap efficiency we will assume a default value of $\epsilon = 0.01$. \change{A brief discussion of the influence of $\mathrm{St}_\mathrm{pbb}$ is shown in Appendix~\ref{sec:pebble_stokes_number}}. The influence of the other parameters in Table \ref{tab:default_set} \change{is} not investigated in this work and they remain constant.



\section{Results}
\label{section:results}

\change{We present our simulation results by first describing the column density evolution of a fixed parameter set and then later discussing the change resulting from parameter variations.}

\subsection{Local evolution}
\label{section:local_evolution}

\begin{figure*}[t]
\centering
\includegraphics[width=\linewidth]{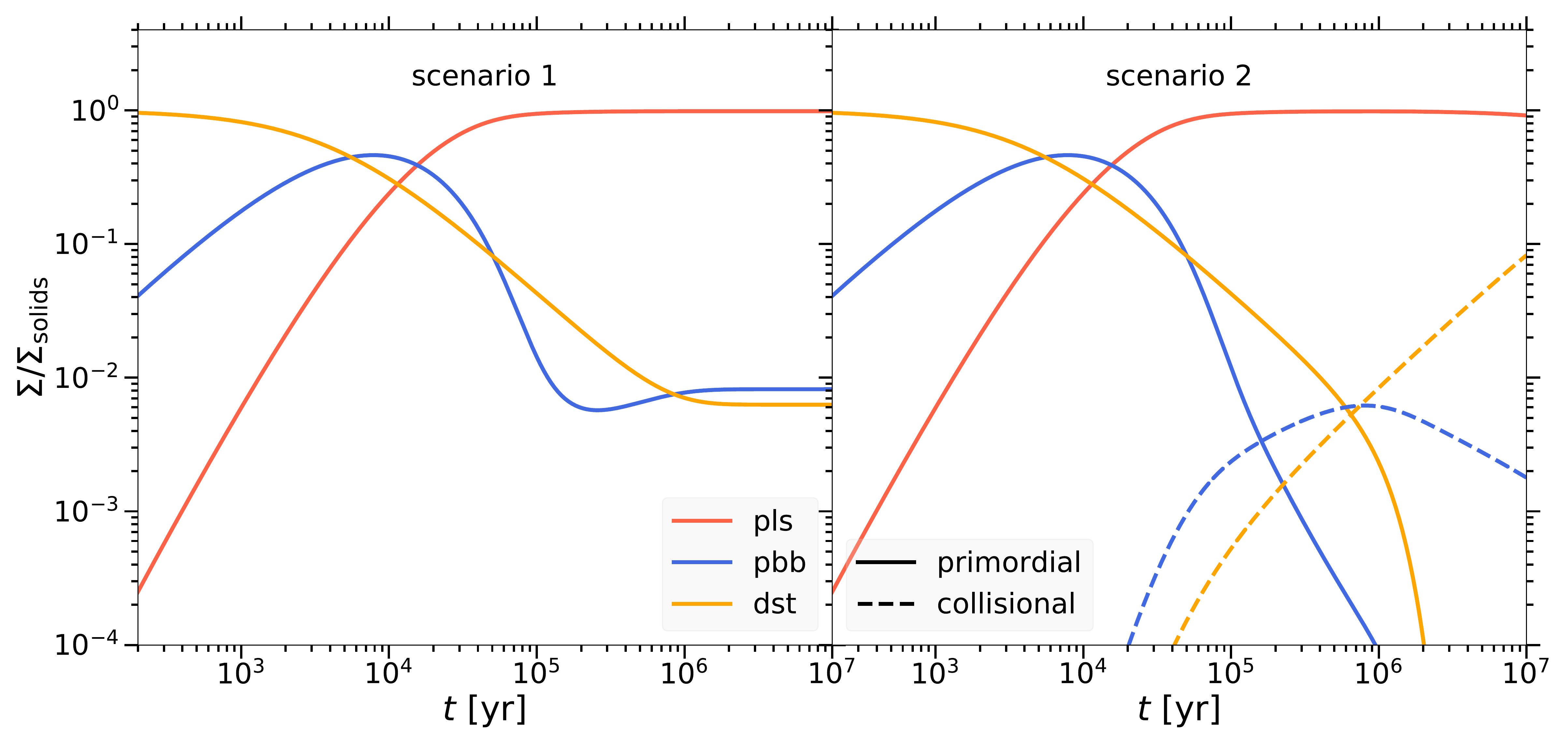}
\caption[]{Local evolution of the different species with column densities normalized by the initial dust column density versus time. The left panel shows the scenario 1 configuration where collisional dust can grow back to (primordial) pebbles, and reach an equilibrium state. The right panel depicts the scenario 2 configuration, where compact collisional dust can not grow pebble sizes. This simulation was done using the default parameter set $R = 10 \mathrm{\ AU}$, $M_\mathrm{disk} = 0.01$ $M_{\odot}$ and $\epsilon = 0.1$.}
\label{fig:evolution_fixed_parameter}
\end{figure*}

\begin{figure*}[t]
\centering
\includegraphics[width=\linewidth]{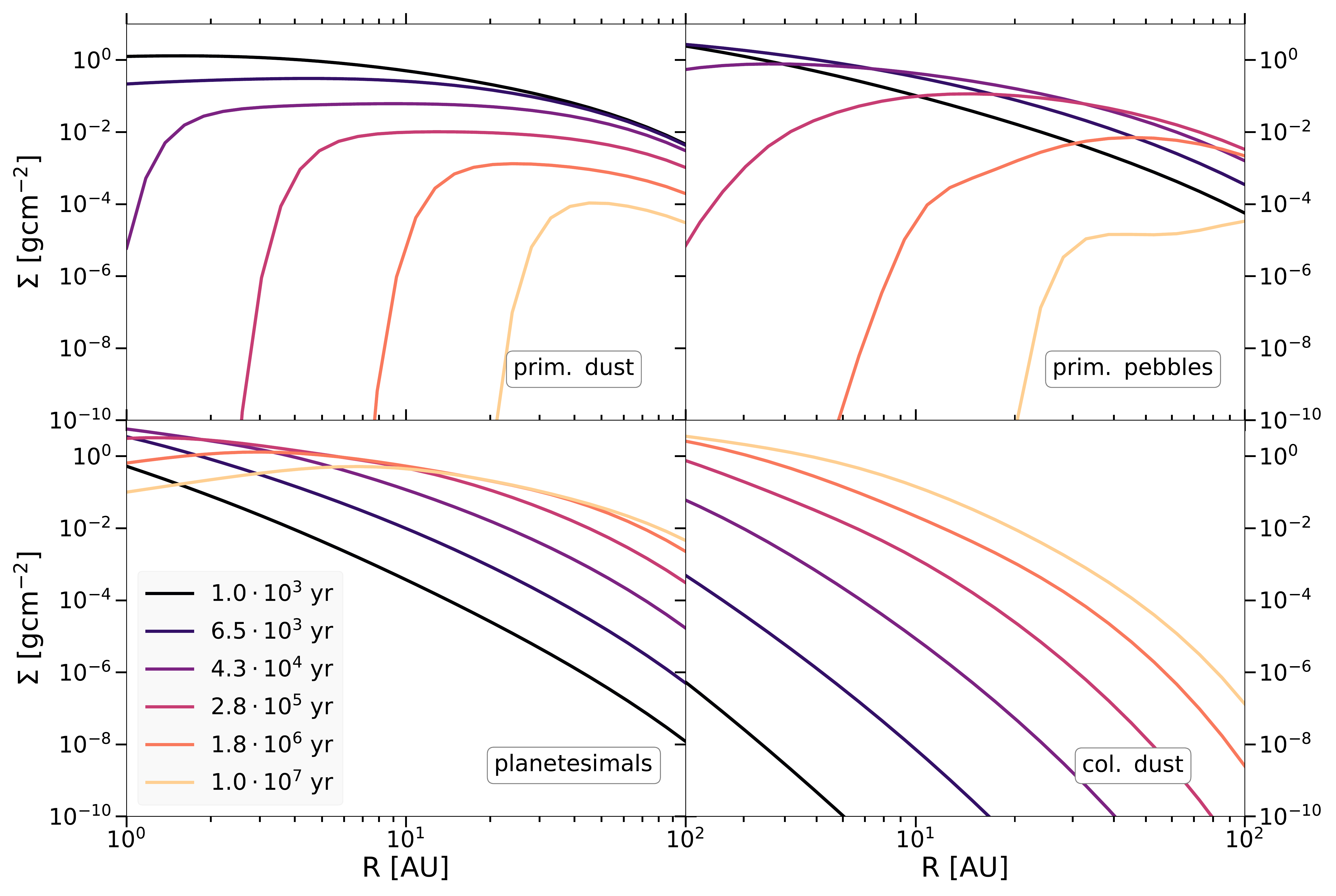}
\caption[]{Column density of different populations is displayed against $R$ for several snapshots. Top left panel corresponds to primordial dust, top right, bottom left and bottom right to primordial pebbles, planetesimals and collisional dust respectively. \change{We depict a combined result of multiple 0-d simulations executed at 20 different radii to allow} an insight in a possible global evolution of the dust profile. \change{All} simulations shown in this figure used $\epsilon = 0.01$.}
\label{fig:sd_vs_R_for_t_default}
\end{figure*}

Figure \ref{fig:evolution_fixed_parameter} displays the local evolution of the normalized column densities $\Sigma_\mathrm{dst,prim}, \Sigma_\mathrm{dst,col}, \Sigma_\mathrm{pbb,prim}, \Sigma_\mathrm{pbb,col}$ and $\Sigma_\mathrm{pls}$. The left panel depicts scenario 1, where the dust-sized fragments of planetesimal collisions grow further to pebble-sizes. This implies the existence of an equilibrium state, which begins after $2\cdot10^6$ yr with this particular set of parameters. It is not surprising that the population of primordial dust follows a monotone decline, since the corresponding differential equation \eqref{eq:primdst} \change{contains only a sink} term. There is a short period during the local evolution where \change{pebble population makes up a significant portion of the column density} --- in Fig. \ref{fig:evolution_fixed_parameter} this is from $4\cdot 10^3$ to approximately $2\cdot 10^4$ yr after the start of the simulation. As we will discuss later, both the duration and the peak height of the pebble column density of this period will vary strongly with the choice of parameters, in particular the trapping efficiency $\epsilon$. The planetesimal population comprises over 98 \change{\%} of the total available column density, i.e. the mass available to the system starting around $5\cdot 10^4$ yr. Collisions between planetesimals naturally will be more common with more planetesimals available leading to the emergence of the two fragment species: collisional pebbles and collisional dust. However, both only make up less than 2 \change{\%} of the total column density once they reach their peak during the equilibrium state.
The right panel in Fig. \ref{fig:evolution_fixed_parameter} shows scenario 2. We see that per definition the collisional dust column density \change{monotonically increases}, since its differential equation \eqref{eq:coldst} only consists of a source term. 

We note that until $\sim 10^6$ yr, the left and right panel are almost congruent with each other. This is expected. The growth of collisional dust, which is the fundamental distinction of scenario 1 and 2, only becomes relevant once planetesimals collisions produce a noteworthy amount of fragments, which for this set of parameters takes about $10^6$ yr.


\subsection{Parameter study}
\label{sectio:parameter_study}

\begin{figure*}[t]
\centering
\includegraphics[width=\linewidth]{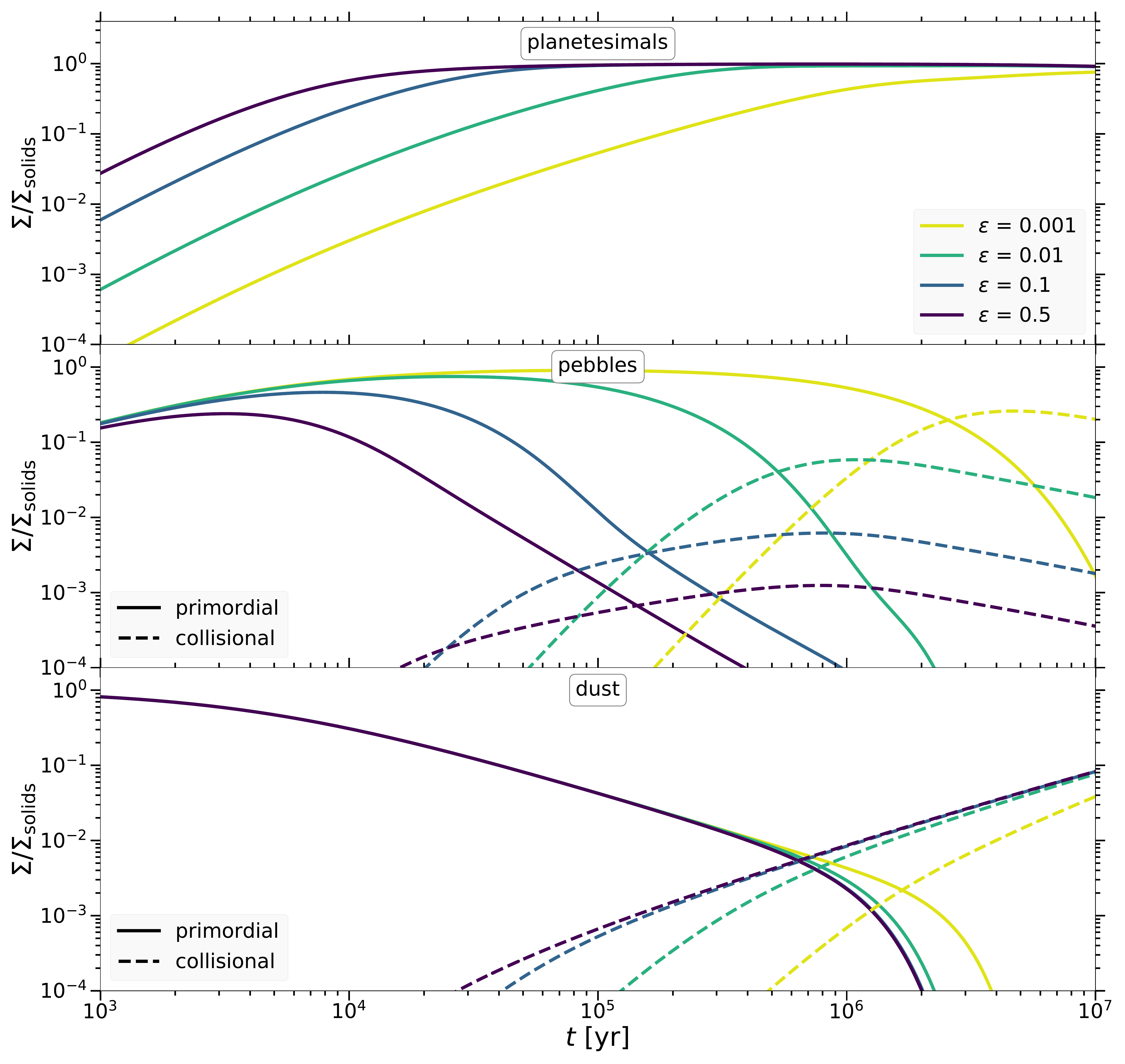}
\caption[]{Local evolution at $R = 10 \mathrm{\ AU}$ for $M_\mathrm{disk} = 0.01 M_{\odot}$ of the normalized column density for different values of the trap efficiency parameter $\epsilon$ (indicated by different colors). The top panel shows the evolution of the planetesimal column density. The \change{middle} panel displays both primordial (solid lines) and collisional pebble populations (dashed lines). Finally, the bottom panel depicts primordial dust (solid lines) and its collisional counterpart (dashed lines). The trap efficiency strongly affects how soon significant planetesimal column densities can be achieved.}
\label{fig:epsilon_variation}
\end{figure*}

\begin{figure*}[t]
\centering
\includegraphics[width=\linewidth]{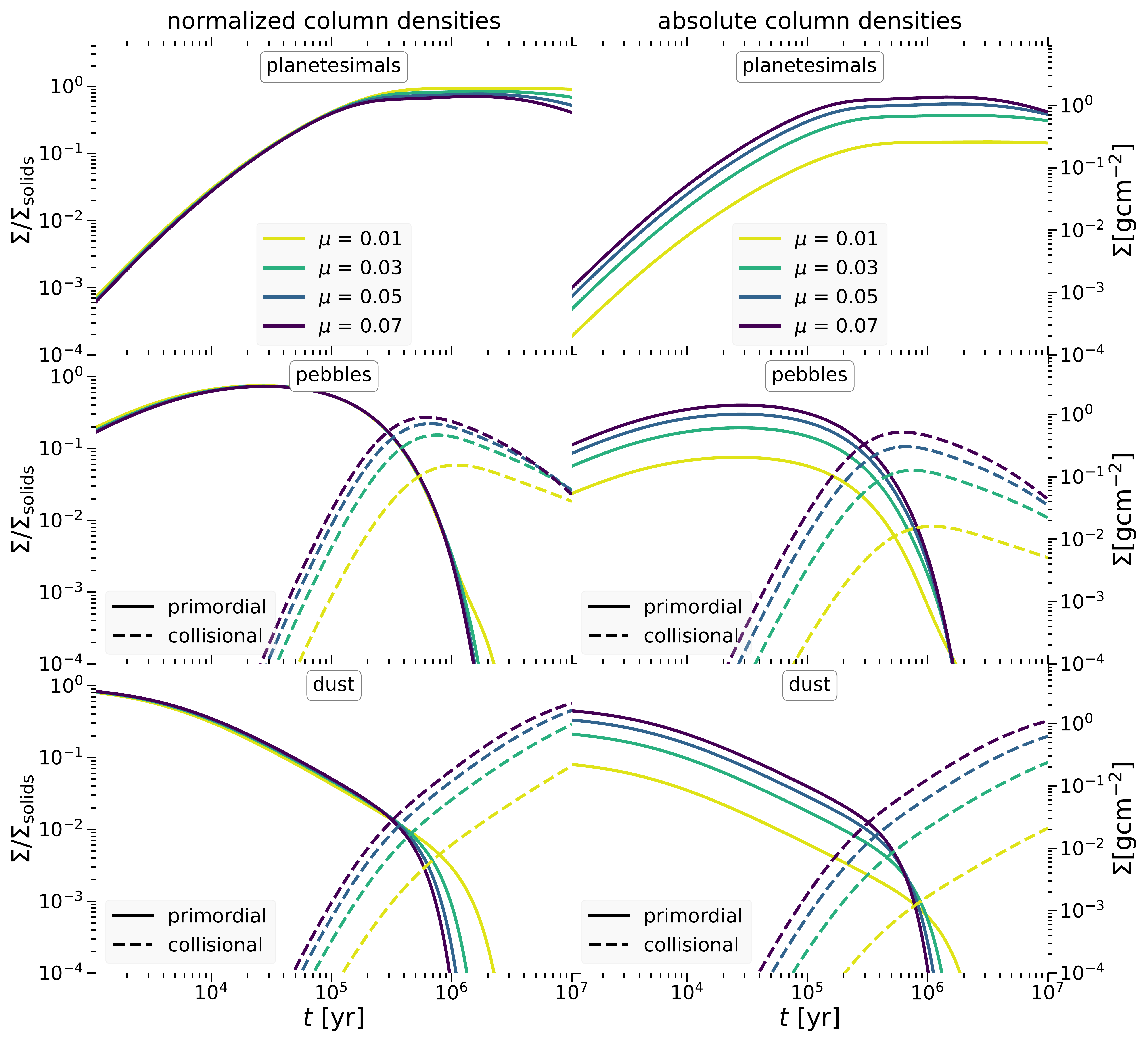}
\caption[]{Local evolution at $R = 10 \mathrm{\ AU}, \epsilon = 0.01$ of normalized (left panels) and absolute (right panels) column density of the species. Panels are arranged in analogy to Fig. \ref{fig:epsilon_variation}: the top panels show the evolution of planetesimal column densities while the middle panels display both primordial (solid lines) and collisional (dashed lines) pebble population. Finally, the bottom panels depict the species of primoridal dust (solid lines) and collisional dust (dashed lines). Different color indicate different values for the disk mass $\mu = M_\mathrm{disk}/M_{\odot} \in [0.01, 0.03, 0.05, 0.07]$. Bright green lines approximately correspond to the MMSN model first discussed in \citet{Weidenschilling1977MMSN}.}
\label{fig:M_variation_sixplots}
\end{figure*}

To develop an understanding of the column density evolution in a system it is crucial to explore various parameters. For this purpose, \newchange{we chose to focus on the}  scenario 2 configuration (see Fig. \ref{fig:flowchart}), \newchange{and base all simulation runs presented in this section on it.}

\subsubsection{\change{Variation of the distance to the star}}

An insight in a possible global evolution can be enabled by executing several simulations for multiple distances to the star $R$. The result of such a simulation run (20 different local simulations at various radii $1 \mathrm{AU} \leq R \leq 100 \mathrm{AU}$ all using $\epsilon = 0.01$) is presented in Fig. \ref{fig:sd_vs_R_for_t_default}, where the column density of the populations is displayed versus $R$ for several snapshots during the simulation. The top left panel shows the primordial dust population, \change{where the decline in time after Eq. \eqref{eq:analytical_solution} is clearly visible}. Moreover, the dust column density declines \change{more rapidly} closer to the star, resulting in an inside-out growth of material \change{\citep[also found in e.g.][]{Birnstiel2012}}. This behavior is expected when one considers the $R$-dependence of the growth timescale in Eq. \eqref{eq:adjustedgrowthtimescale}, \change{which is much shorter for small $R$} than in the outer disk.

The same effect can be seen upon inspection of the top right panel in Fig. \ref{fig:sd_vs_R_for_t_default}, which displays the primordial pebble column density. \change{In the beginning at $10^3$ yr,} the pebble population increases in the inner disk much faster than in the outer disk resulting in a much steeper profile compared to the primordial dust population. Further, we can identify that removal of primordial pebbles via planetesimal formation likewise is an inside-out process\change{, as a direct implication of inside-out dust growth.}

\change{The planet\newchange{e}simal population naturally mirrors this behavior too, as} pictured in the bottom left panel of Fig. \ref{fig:sd_vs_R_for_t_default}: \change{until $2.5 \cdot 10^5$ yr}, the planetesimal column density increases more rapidly at smaller radii. However, once the local pebble supply is depleted, planetesimal formation slows down drastically and collisions start to decrease the \change{planetesimal} local column density again. This also happens \change{first} at smaller radii, because here, the pebble supply is depleted first. Therefore, the peak of the planetesimal column density also moves towards the outer disk with increasing time\change{, resulting in an inside-out formation of planetesimals as also shown by \citet{Lenz2019}.} \change{The same cannot be argued for further growth to planets. Here, in addition to the radial planetesimal mass distribution, dynamical stirring and the available pebble supply become relevant, introducing complex $R$-dependencies.}

Finally, the bottom right panel of Fig. \ref{fig:sd_vs_R_for_t_default} displays the column density of collisional dust particles, which will eventually make up the bulk of the column density as already seen in the right panel of Fig. \ref{fig:evolution_fixed_parameter}. \change{For a fixed $R$, it increases all the time}, again \change{first} in the inner disk \change{and later} in the outer disk, where it never surpasses the planetesimal column density.  

\subsubsection{\change{Variation of the efficiency of planetesimal formation}}

Further, we investigate the influence of the trap efficiency $\epsilon$ on the column density evolution. All other parameters are kept constant according to the default parameter set. The trap efficiency is per definition $0\leq \epsilon \leq 1$, with $\epsilon = 1$ implying that the entire incoming pebble flux is trapped and converted into planetesimals and $\epsilon = 0$ meaning that the trap is not capturing and forming planetesimals.

A comparison of the local column density evolution for different $\epsilon$-values ($\epsilon$ = 0.001, 0.01, 0.1, 0.5) is shown in Fig. \ref{fig:epsilon_variation}. 
The top panel depicts the evolution of the planetesimal column density. For all values of $\epsilon$ the general trend already seen in the right panel of Fig. \ref{fig:evolution_fixed_parameter} is replicated. 
The column density of planetesimals increases until they make up the vast majority of the column density available in the system. However, the \change{rate} of planetesimal formation strongly depends on $\epsilon$. 
During the initial phase of planetesimal formation (until $10^4$ to $10^6$ yr depending on the value of $\epsilon$) the planetesimal column density \change{is linear in $\epsilon$ after Eq. \eqref{eq:formationrate}}. However, this linearity ceases to exist once the normalized column density approaches unity.
 
On the top panel of Fig. \ref{fig:epsilon_variation}, the curves for the different $\epsilon$-values seem to roughly coincide once the decline of the planetesimal population sets in. This is not surprising, since the collision rate in Eq. \eqref{eq:collisionrate} does not depend directly on $\epsilon$. \change{Here, planetesimal formation is insignificant, because there is only few pebbles available.} \change{We point out, that} for all $\epsilon$-values almost all the mass ends up within the planetesimal population --- \change{for $\epsilon = 0.5,0.1,0.01,0.001$ starting at approximately $2\cdot 10^4, 5 \cdot 10^4, 6 \cdot 10^5 $ and $10^7$ yr respectively.}

The \change{bottom} panel in fig.\ref{fig:epsilon_variation} displays the evolution of primordial dust (solid lines) as well as collisional dust (dashed lines) for the same four values of the trap efficiency parameter $\epsilon$. In the \change{middle} panel, one can see an equivalent plot for primordial pebbles (solid lines) and collisional pebbles (dashed lines).
The influence of $\epsilon$ on the evolution of primordial dust is negligible, since the dust growth timescale does not directly depend on $\epsilon$, \change{as seen in} Eq. \eqref{eq:adjustedgrowthtimescale}. 
However, both the peak height and width, as well as the rate of decrease of the primordial pebble population strongly depend on $\epsilon$ as shown in the plot. Inefficient traps leave behind a significant population of primordial pebbles, while more efficient traps can drain the pebble supply more rapidly. 
The collisional pebble population peaks once production of collisional pebbles can not keep up with drainage of pebbles through planetesimal formation, i.e. in Eq. \eqref{eq:colpbbdot} $\dot{\Sigma}_\mathrm{pbb, col} = 0$. This is strongly influenced by the column density available in primordial pebbles, since they outnumber their collisional counterparts at early times. 
The intersection with the \change{horizontal} axis of both the collisional dust and the collisional pebbles curve depends on $\epsilon$, because $\epsilon$ is what determines when enough planetesimals can be formed for collisions to produce noteworthy amounts of fragments (see top panel of Fig. \ref{fig:epsilon_variation}).

\subsubsection{\change{Variation of the initial disk mass}}

Protoplanetary disks are observed with a number of different masses. \citet{Andrews2010} found disk masses in the range of $M_\mathrm{disk} = 0.004 - 0.143 M_{\odot}$ in the $\sim$ 1 Myr old Ophiuchus star-forming region. The MMSN model discussed in \citet{Weidenschilling1977MMSN} describes a disk of solar composition containing the minimum amount of solids necessary to form the eight planets and the asteroid belts of today's solar system, by only considering their masses and today's positions. \change{In the MMSN model}, rocky and icy objects have a total mass of roughly $2 \cdot 10^{-4} M_{\odot}$ while gas has a total mass of $1.3 \cdot 10^{-2} M_{\odot}$. \newchange{Although} effects such mass loss via photoevaporation are not considered, it is still expedient to execute parameter variations of the disk mass. 

Such a parameter study is shown in Fig. \ref{fig:M_variation_sixplots}. We define a dimensionless disk mass 
\begin{equation}
\mu := \frac{M_\mathrm{disk}}{M_{\odot}}
\end{equation}
and investigate values ranging from $\mu = 0.01$, approximately corresponding to the MMSN after \citet{Weidenschilling1977MMSN} up to $\mu = 0.07$. More massive disks \change{are expected to be gravitationally unstable} and are additionally \change{uncommon} when taking recent observations by \citet{Andrews2009} and \citet{Andrews2010} into account. The other model parameters are kept constant. The normalized column density is depicted in the left panels whereas the right panels show its absolute value. Panels are arranged in analogy to Fig. \ref{fig:epsilon_variation}: the top panels show the evolution of the planetesimal column densities. The maximum absolute planetesimal column density naturally \change{increases} with disk mass, since a higher disk mass implies more available material to form planetesimals. Interestingly this trend is reversed when looking at the normalized planetesimal column density evolution in the top left panel of Fig. \ref{fig:M_variation_sixplots}. Less massive disks \change{have a lower relative planetesimal fraction} than more massive disks\change{, because the effect of planetesimal collisions becomes more important with increasing disk mass.}

The middle panels of Fig. \ref{fig:M_variation_sixplots} display the primordial (solid lines) and the collisional pebble population (dashed lines). In the bottom panels, one can see the primordial (solid lines) and collisional (dashed lines) dust populations. The absolute primordial dust and primordial pebble content increases with increasing disk mass. However, there seems to be no strong correlation between normalized primordial column density and disk mass (middle left and bottom left panels of Fig. \ref{fig:M_variation_sixplots}). Only starting at $\sim 5 \cdot 10^5$ yr, the panel shows that the primordial dust supply is drained faster for more massive disks. The peak height of the absolute collisional pebble column density in the middle right panel correlates strongly with disk mass. This trend is replicated for the normalized column density. For the most massive disks pebbles make up around 30\% of the total available column density at the peak of the pebble population (roughly $4 \cdot 10^5$ yr). For MMSN-like disks, the peak height is not only lower ($\sim$ 2 \% of available column density), it also occurs slightly later during the evolution ($10^6$ yr). Similarly, both absolute and normalized collisional dust column density evolution correlate with disk mass. For massive disks the fraction of collisional dust is generally larger than for less massive disks and also starts to increase somewhat sooner ($5 \cdot 10^4$ yr for the most massive disk compared to $10^5$ yr for the MMSN disk) and with a steeper slope.

In \change{our} model, evolution of dust, pebbles and planetesimals is independent of the turbulence parameter $\alpha$, \change{but only valid for disks where turbulence is strong enough to set relative dust velocities as discussed in Sect. \ref{section:dust_growth}}. Further and more importantly, the pebble Stokes number is for simplicity kept constant at $\mathrm{St}_\mathrm{pbb} = 0.1$ \change{for our parameter study. We discuss implications in \ref{sec:pebble_stokes_number}}. A more complex model would introduce a dependence on $\alpha$ as seen in \citet{Birnstiel2009}. Additionally, turbulence influences radial transport of disk material strongly, which is also neglected in this simple model. Finally, in this paper, we will refrain from performing parameter studies with $\epsilon_{\mathrm{dg}}^0$ and $r_\mathrm{pls}$ since their influence on the column density evolution is not the main focus of this work.


\subsection{Estimating the evolution of the mass distribution}
\label{section:mass_distribution}

\begin{figure*}[t]
\centering
\includegraphics[width=\linewidth]{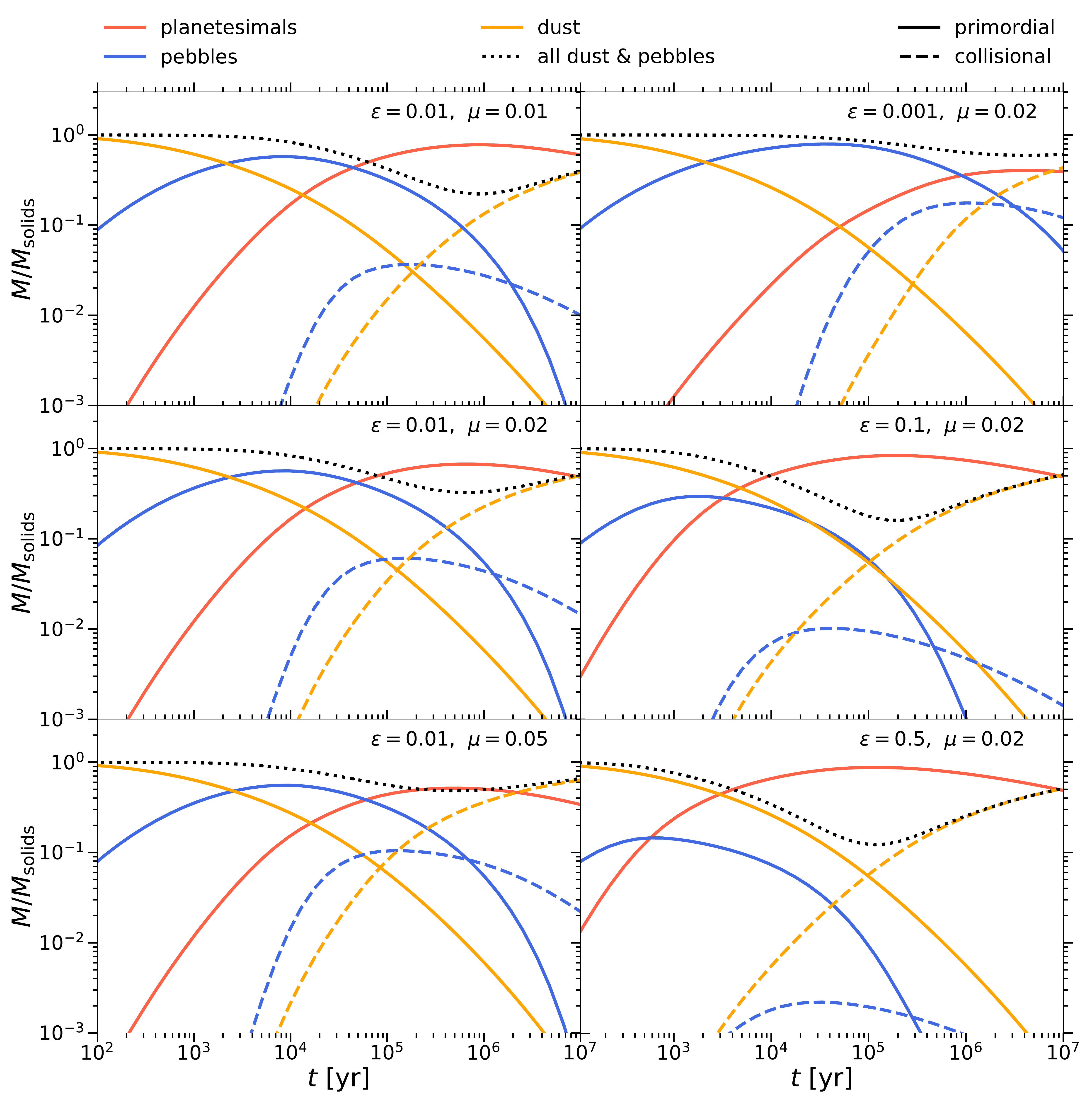}
\caption[]{Evolution of the normalized total mass of solid particles in the disk for different values of $\epsilon \in \{0.001, 0.01, 0.1, 0.5\}$ and disk mass $\mu = M_\mathrm{disk}/M_{\odot} \in \{0.01, 0.02, 0.05\}$. Different colors correspond to the different species (red: planetesimals, yellow: dust, blue: pebbles, black \change{dotted} lines: dust and pebbles combined)\change{, the solid lines to primordial and the dashed lines to collisional populations.} In the left panels, disk mass increases from top to bottom while the trap efficiency $\epsilon$ remains constant. Likewise, in the right panels, trap efficiency increases downwards, however the middle left panel would also fit within this sequence between the top right and the middle right panel.}
\label{fig:mass_evolution_different_eps_and_mus}
\end{figure*}

\begin{figure}[t]
\centering
\includegraphics[width=\linewidth]{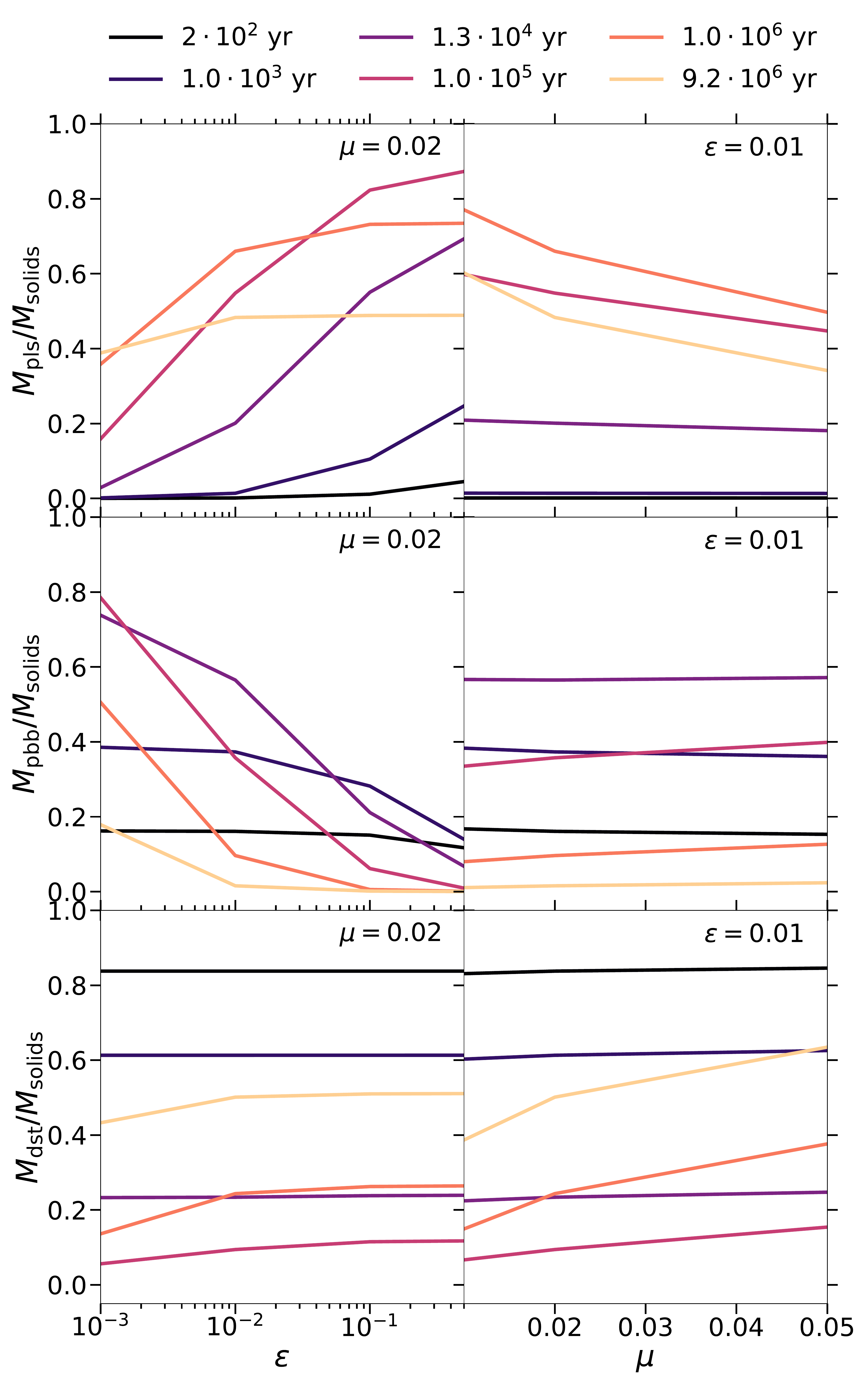}
\caption[]{Mass fraction in planetesimals (top panels), pebbles (middle panels) and dust (bottom panels) vs trap efficiency $\epsilon$ (left panels) and disk mass $\mu = M_\mathrm{disk}/M_{\odot}$ (right panels). In the left panels, the disk mass is fixed at $\mu = 0.02$ while the trap efficiency is fixed to $\epsilon = 0.01$ in the right panels. Colors correspond to different times in the disk evolution. Data is taken from the same simulation runs that were depicted in Fig. \ref{fig:mass_evolution_different_eps_and_mus}. \change{However, Fig. \ref{fig:mass_vs_eps_and_mu} does not differentiate between primordial and collisional populations and depicts their sum.}}
\label{fig:mass_vs_eps_and_mu}
\end{figure}

The evolution of the mass that is contained within planetesimals is of particular interest since \change{planetesimals are too cold to be observed in the infrared and they also} provide too little column density to be detected via \change{scattered light} observations. Hence, this section presents an approximation of the evolution of the mass distribution. The total mass $M_i$ \change{of a given population $i$ that is contained within $R_1$ and $R_N$ } at a certain time $t$ \change{is given by integrating the column density at $t$ over $r$
\begin{equation}
\label{eq:M_approx_1}
M_i(R_1\leq r\leq R_N,t) = 2 \pi \int_{R_1}^{R_N} r \Sigma_i(r,t) \mathrm{d}r.
\end{equation}
We can approximate this integral by} performing multiple simulation runs \change{$k$} at different radii \change{$R_k$} and then \change{summing over them, i.e., using the trapezoidal rule,
\begin{align}
\nonumber
M_i(R_1\leq r\leq R_N,t) \approx&\, \pi \sum_{k = 1}^{N-1} \left(R_{k+1} - R_{k}\right)
\\
&\times\left[R_{k}\Sigma_i(R_k,t) + R_{k+1}\Sigma_i(R_{k+1},t)\right].
\end{align}
}
\change{We set $N=20$, $R_1 = 1 \mathrm{\ AU}$, and $R_{20} = 100 \mathrm{\ AU}$. For radii larger than $100$ AU, the column density profile in Eq. \eqref{eq:initial_dust_profile} declines exponentially and prevails the $r^2$-dependence in Eq. \eqref{eq:M_approx_1}, rendering the contribution to the total mass beyond this point insignificant.} \change{We point out that} this procedure is only expected to deliver a rough estimate of the total mass evolution in the disk. For a more precise picture, spatial mass transport certainly has to be considered.

Figure \ref{fig:mass_evolution_different_eps_and_mus} displays the evolution of the normalized mass of solid particles in the disk, as obtained via the algorithm explained above. Different colors correspond to different species (yellow for dust, blue for pebbles and red for the planetesimal population). Black dashed lines represent the sum of \change{all} dust and pebble particles, since we can not expect to be able to distinguish these species with observations. Different panels originate from simulation runs with different values for trap efficiency  
($\epsilon \in \{0.001, 0.01, 0.1, 0.5\}$) and disk mass  $\mu = M_\mathrm{disk}/M_{\odot}$ ($\mu \in \{0.01, 0.02, 0.05\}$, where the lower boundary roughly approximates the MMSN-mass from \citet{Weidenschilling1977MMSN}), which is indicated in the top right corner of each panel. We note that while $\mu$ represents the total disk mass containing both solids and gas, the \change{vertical axis} in each panel of Fig. \ref{fig:mass_evolution_different_eps_and_mus} only displays the mass of each of the solid species. The data is normalized with respect to the total mass of solid particles ( $\epsilon_{\mathrm{dg}}M_\mathrm{disk}$). The mass evolution \change{for} each panel \change{is comparable to the} column density \change{evolution} in Fig. \ref{fig:evolution_fixed_parameter}. In the beginning, all mass in the system is contained within the combined dust population. After a period during which the combined pebble population makes up a non-negligible part of the total mass, the planetesimal population starts to dominate. Once the collisions take over, the mass \change{is being transferred} back to the combined dust population. When comparing the three panels on the right side as well as the middle left panel, each corresponding to different $\epsilon$-values for a constant disk mass of $\mu = 0.02$, one discovers, that $\epsilon$ both influences the time \change{where the maximal planetesimal mass is reached and its peak value}. Similar to Fig. \ref{fig:epsilon_variation}, a higher $\epsilon$-value implies that the planetesimal population peaks at earlier times and with a greater maximal peak height. In the top right plot, where the trap efficiency parameter was set to the low value of $\epsilon = 0.001$, the planetesimal population (red line) \change{never becomes more massive} than combined dust-and-pebble population (green line). It's peak is roughly equivalent to a mass of 26.8 $M_\mathrm{Earth} $. A tenfold increase of $\epsilon$ implies approximately a 20 \% increase of the maximal planetesimal mass and this peak occurring roughly ten times earlier. In the left panels, disk mass \change{increases from the top to the bottom panel} while trap efficiency $\epsilon$ remains constant at $\epsilon = 0.01$. As expected from Fig. \ref{fig:M_variation_sixplots}, more massive disks \change{tend to have a smaller relative planetesimal fraction} than less massive disks, leading to a lower ratio of planetesimal mass to dust-and-pebble combined mass for more massive disks. An interesting observation is that in the top left panel, the maximal planetesimal mass is also approximately 26.8 $M_\mathrm{Earth}$ similar to the top right panel, implying a doubled normalized disk mass $\mu$ would be approximately balanced by a tenfold decrease of $\epsilon$. When observing these model disks at the times of the respective planetesimal mass peak, one would conclude \change{significantly} different dust masses, even though the planetesimal mass is comparable. The evolution of the dust mass in the top left panel would \change{for example} compare better to the bottom right panel \change{with} $\epsilon = 0.5$, \change{where planetesimals reach the mass fraction peak earlier than in the top left panel.}.

In Fig. \ref{fig:mass_vs_eps_and_mu}, we further visualize how the mass in solids is distributed among the populations and how this depends on the efficiency $\epsilon$ (left panels) and the dimensionless disk mass $\mu$ (right panels). Data are taken from the same simulation runs as Fig. \ref{fig:mass_evolution_different_eps_and_mus}, while colors indicate the time the snapshot was taken, i.e. the disk age. The evolution of the dust content (bottom panels) does only depend on $\epsilon$ and $\mu$ for times $\gtrsim10^5$ yr. This is because only then planetesimal collisions start to produce a significant amount of dust. The amount of these fragments is determined by the planetesimal column density, which depends on $\epsilon$ and $\mu$ as seen in the top panels of Fig. \ref{fig:mass_vs_eps_and_mu}. Here, while the mass in planetesimals always increases with $\epsilon$, the relation is clearly non-linear. This is explained \change{due to the fact}, that the pebble supply is limited and therefore the maximum planetesimal column density is independent of $\epsilon$, as seen Fig. \ref{fig:epsilon_variation}. In the top right panel of Fig. \ref{fig:mass_vs_eps_and_mu}, we can again observe that the planetesimal fraction is indeed decreasing with increasing disk mass, i.e. less massive disks are more efficient in forming planetesimals, which is in accordance to Fig. \ref{fig:M_variation_sixplots}. The behavior of the pebble fraction is depicted in the \change{middle} panels of Fig. \ref{fig:mass_vs_eps_and_mu}. \change{It is approximately constant in $\mu$, because the initial growth phase from dust to pebbles only depends on the dust-to-gas ratio, which is picked the same for all $\mu$. The small differences originate from the fact that the significance of planetesimal collisions depends on total mass.} \change{However, the pebble fraction} correlates strongly with $\epsilon$ (middle left panel), because more efficient planetesimal formation implies that the pebble supply is drained faster. We also observe that during \change{late} times \newchange{at $t = 9.2 \cdot 10^6$~yr}, the mass in pebbles \newchange{is rather} independent of $\epsilon$. \newchange{This} is due to planetesimal formation being insignifcant in comparison to planetesimal collisions, which do not depend on $\epsilon$.



\section{Limitations and advantages of the model}
\label{section:limitations}
Our model is limited by various factors. Most strikingly, its locality and the resulting absence of spatial transport of material constitutes a strong limitation, as our model can not currently cover scenarios where the pebble drift timescale is shorter than the conversion timescale for pebbles into planetesimals. The 0-dimensional model can also not consider any effects proceeding perpendicular to the disk plane, such as dust settling \change{and vertical turbulent stirring}. The dependence of the model on disk turbulence was eliminated by assuming that particle growth does not scale with disk turbulence, i.e. $\mathrm{St} \gg \alpha$. \change{Additionally}, viscous gas evolution and photoevaporation \citep[see e.g.][]{ Ercolano2009, Owen2011, Nakatani2018, Picogna2019} were also not considered. In comparison to \citet{Birnstiel2012}, who derive the maximum pebble Stokes number by taking the dust growth limiting effects drift and fragmentation into account, we fix the pebble Stokes number to a guiding value at $\mathrm{St}_\mathrm{pbb} = 0.1$ during the entire disk evolution. \change{The influence $\mathrm{St}_\mathrm{pbb}$ is discussed in Appendix~\ref{sec:pebble_stokes_number}}. This could be improved by implementing these growth barriers, which depend on gas and particle density. This would allow for a time- and space-dependent pebble Stokes number as in \citet{Birnstiel2012}. \change{Alternatively, dust coagulation could be modelled in more detail by implementing a particle size grid and solving the Smoluchowski equation \citep{Smoluchowski1916}, as was done by \citet{Lenz2019}, who we further compare our results to in Appendix~\ref{sec:comparison_to_lenzetal2019}.}

Many observed \change{circumstellar} disks contain substructures, mainly rings as first seen by \citet{ALMA2015} in HL Tauri or recently in \citet{Andrews2018} presenting results from the \textit{Disk Substructure at High Angular Resolution Project}. As the underlying physical process is still in debate, as recently discussed in \citet{Marel2018}, disk substructures are not considered in this work's model for gas column density evolution.  Furthermore, ice lines, which mark the border where condensation of volatiles such as e.g. water is possible, were not considered. The Clausius–Clapeyron \citep{Clausius1850} relation describes the slope of the tangent dividing the two phases in a pressure-temperature diagram, which for typical concentrations of H$_2$O would correspond to approximately $T_\mathrm{g} \gtrsim $ 150 K -- 200 K. The addition of the water ice line would imply a significant kink in the dust profile as seen for example in the model by \citet{Drazkowska2017}. \change{In their work, ice lines mark a favorable location for planetesimal formation.} Further not considered is the accretion of solids onto the central star, removing material in the inner disk and decelerating or even preventing planetesimal formation in these regions.

Chemical composition of solids and how it depends on $R$ is also not part of this model. One may expect the composition of a dust particle to influence its growth \change{growth rate. Additionally, the} composition of a planetesimal \change{affects the size distribution of collisional fragments \citep{Johnson2012}}

The fact that the local model does not consider different planetesimal sizes, is also a restriction, as the outcome of a collision is dependant on the mass and sizes of the colliding planetesimals. \change{The temporal evolution of the planetesimal size distribution is not considered. Further, we neglect any planet-forming processes such as} pebble accretion \newchange{\citep[see e.g.][]{Ormel2010, Ormel2017,Rosenthal2018, Lambrechts2019}, which may} drain the pebble supply and therefore decelerating the birth of new planetesimals. At planetesimal sizes of $\sim$ 100 km, the efficiency of pebble accretion is minimal. \newchange{Therefore, our assumption of a fixed planetesimal size of $\sim$ 100 km does not allow for pebble accretion to be efficient.} \change{Likewise, planetesimal accretion \citep[e.g.][]{Kokubo2012} is also neglected.}

Lastly, various limiting assumptions go into this work's model for planetesimal formation, as we condense unknown information about the physical nature of particle traps in the trapping efficiency parameter $\epsilon$. \change{Besides the streaming instability \citep{Youdin2005}, other hydrodynamical processes such as subcritical baroclinic instability \citep{Klahr2003}, convective overstability \citep{Klahr2014}, or vertical shear instability \citep{Urpin1998} may also affect the pebble trapping mechanism \citep{Klahr2018}}. While \citet{ Pfeil2019} show that hydrodynamic instabilities can act throughout the entire disk, the one that dominates the formation of traps may depend on the location in the disk. Therefore, it is plausible that $\epsilon$ may also depend on $R$, as well as on the pebble Stokes number.

Strength of our local model is, that it considers three defining processes in planetesimal formation and evolution theory and integrates them into one simple mode, which is both easy to understand and due to its locality easy to implement as well as computational inexpensive. \newchange{Our model is able to form significant column densities of planetesimals everywhere in the disk and fast ($\sim 10^6$~yr), i.e. well before the onset of gas dissipation. Therefore, our toy model can be used to inform initial conditions of late-stage protoplanetary disk evolution models.}


\section{Summary and conclusions}
\label{section:summary}

We investigated a local model of \change{pebble flux-regulated} planetesimal formation \change{\citep{Lenz2019}}, dust growth \change{\citep{Birnstiel2012}} and planetesimal collisions to study the evolution of the planetesimal to dust and pebble ratio using an enclosed set of coupled differential equations. Aim of the performed parameter study was to develop an understanding of the potentially existing mass in the planetesimal population for a certain amount of dust and pebbles and how it relates to our model parameters.

In Sect. \ref{section:principles}, we discussed the principles on which our local model is based on. To each of the three processes dust growth, planetesimal formation, and planetesimal collisions a timescale is attached to, which allows the formulation of rate equations (see Eqns. \eqref{eq:firstrateeq} to \eqref{eq:coldst}) describing the evolution of the enclosed system. To solve this set of differential equations a code following a one-step procedure was written. To keep the code simple and numerically inexpensive,  radial transport of material and gas evolution was neglected, besides planetesimals, only two small particle populations --- dust and pebbles --- were considered, and the Stokes number of the latter was assumed to stay constant at $\mathrm{St}_\mathrm{pbb} = 0.1$, which is roughly the maximum Stokes number in \citet{Birnstiel2012}. \change{Implications are discussed in Appendix \ref{sec:pebble_stokes_number}}. The small particle populations were further differentiated into primordial/ pristine particles and fragments from planetesimal collisions. We discussed two scenarios. In the first, collisional dust was allowed to grow back to pebbles and form planetesimals again and in the second this was forbidden. In the former scenario, this leads to a steady state with constant column densities as visualized in the left panel of Fig. \ref{fig:evolution_fixed_parameter}. Here, column densities remain constant. However, for our parameter study we used the latter scenario, which is the more realistic approach, because fragments of planetesimal collisions are assumed to be too compact to undergo growth via sticking. Here, the local evolution can be broken down into three phases:
\begin{enumerate}
\item During the first stage of the disk evolution, primordial dust grows to pebble sizes, the dust population shrinks to one hundredth of its initial value in roughly $10^5$ yr. For the behavior of primordial dust the simple analytical solution in Eq. \eqref{eq:analytical_solution} can be found.
\item The phase in which planetesimals are the dominating species naturally follows the increase of pebbles that are available for planetesimal formation. \newchange{This occurs fast, i.e. within $\sim 10^6$ yr, and everywhere in the disk.} As a feedback, the pebble population is drained again, limiting planetesimal formation. At the same time collisions between planetesimals produce collisional fragments in both pebble and dust sizes. While the former species can form new planetesimals and prolong the planetesimal dominated phase, the collisional dust species starts to steadily increase.
\item Once the pebble supply is depleted, the planetesimal population shrinks and dusty fragments \newchange{start to become} the dominating species in the system. This occurs at approximately $10^7$~yr. \newchange{Beyond $10^7$~yr gas dissipation cannot be neglected \citep{Hernandez2007, Mamajek2009, Fedele2010, Pfalzner2014} and our model assumptions fail.}
\end{enumerate}
In our parameter study, we focused on the effects of distance to the star $R$, trap efficiency $\epsilon$ and disk mass $M_\mathrm{disk}$. Note again, that $\epsilon$ incorporates both the efficiency of the pebble trapping mechanism as well as the efficiency of planetesimal formation itself. \change{We summarize our parameter study below.}
\begin{itemize}
    \item The dust growth timescale is much shorter close to the star than in outer regions, which has the \change{direct} implication that the entire local evolution \change{is occuring on shorter timescales} for small $R$ than for larger $R$ as seen in Fig. \ref{fig:sd_vs_R_for_t_default}. \change{This inside-out growth of dust is in line with e.g. \citet{Brauer2008, Birnstiel2010, Birnstiel2012} \newchange{or} \citet{Krijt2016}. Inside-out formation of planetesimals agrees with \citet{Lenz2019}.} 
    \item High $\epsilon$-values generally imply an acceleration of the planetesimal formation process, leading to a longer planetesimal dominated phase (see Fig. \ref{fig:epsilon_variation}).
    \item When studying different disk masses in \change{as shown in} Fig. \ref{fig:M_variation_sixplots}, we found that while more massive disks have a higher absolute planetesimal column density than less massive disks, their normalized planetesimal column density is lower, \change{because planetesimal collisions are more significant in massive disks compared to less massive disks.}
\end{itemize} 
Lastly, we showed an estimate for the mass evolution of the disk in Fig. \ref{fig:mass_evolution_different_eps_and_mus} and Fig. \ref{fig:mass_vs_eps_and_mu}. Here, we display\newchange{ed} how the distribution of material among the three populations evolves in time and depends on planetesimal formation efficiency and disk mass. Therefore, our local toy model can be a potential tool for constraining the material in planetesimals in observed protoplanetary disks.

By investigating the influence of $\epsilon$ and $M_\mathrm{disk}$ on the mass evolution in Fig. \ref{fig:mass_evolution_different_eps_and_mus}, we are able to make statements on the potentially existing mass in planetesimals for a given amount of pebbles and dust. Provided the total mass of a certain disk is known and measured independent from dust mass by observations \change{such as e.g. \citet{Pascucci2016},} or modeling, the more dust is observed the less mass can be expected to be compromised in planetesimals relative to the total disk mass. If one can determine the age of the protoplanetary disk, e.g. via the position of the central star (or its neighbors) in the Hertzsprung-Russel diagram, and provided both total disk mass and dust portion relative to it are known \change{from empirical data}, the toy model may constrain the value of the trap efficiency $\epsilon$ as seen in Fig. \ref{fig:mass_evolution_different_eps_and_mus} and Fig. \ref{fig:mass_vs_eps_and_mu}. If one, however, is unsure about the total mass of a certain disk, which work by \citet{Andrews2010} suggest is typically the case, it becomes challenging to predict the planetesimal mass for a given observed dust mass, since the dust and pebble column density evolution can look almost identical for two different sets of trap efficiency $\epsilon$ and disk mass $M_\mathrm{disk}$.

Despite many simplifications made in our model, it generated results connecting important model parameters like trap efficiency $\epsilon$ and disk mass $M_\mathrm{disk}$ to the local column density evolution of three different species dust, pebbles and planetesimals. Additonally, it helped us to further our understanding for the evolution of the mass distribution over the course of over \newchange{$10^7$}~yr ranging from the birth of the protoplanetary disk until its \newchange{dissipation}.

\begin{acknowledgements}
K.G. thanks Ruth A. Murray-Clay and U.C. Santa Cruz for hosting him as a graduate visitor. C. L. acknowledges funding from the Deutsche Forschungsgemeinschaft (DFG,  German  Research  
Foundation) as part of the Schwerpunktprogramm (SPP, Priority Program) SPP 1833 ``Building a Habitable Earth''. Finally, the authors are grateful to the anonymous referee for an insightful and detailed report which greatly helped to improve this paper.
\end{acknowledgements}


\bibliographystyle{aa.bst} 
\bibliography{bibliography_paper} 



\appendix

\section{Comparison to the planetesimal synthesis model in \citet{Lenz2019}}
\label{sec:comparison_to_lenzetal2019}

\begin{figure*}[t]
\centering
\includegraphics[width=0.8\textwidth]{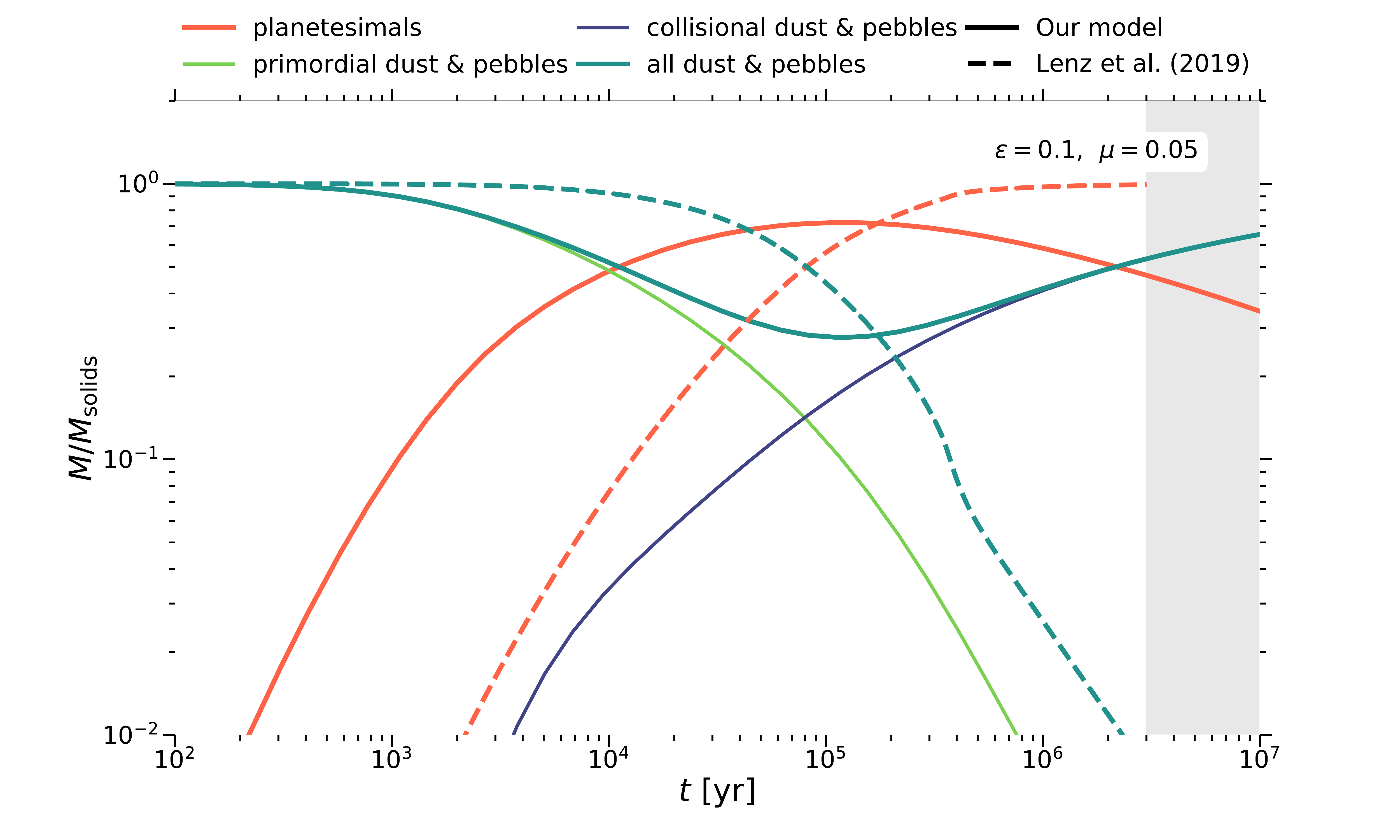}
\caption[]{\change{Evolution of the normalized total mass of solid particles in the disk for our model (solid lines) and the model in \citet{Lenz2019} (dashed lines) for $\alpha = 10^{-3}$. Both models used $\epsilon = 0.1$, $\mu = 0.05$, $R_\mathrm{C} = 35\,\mathrm{AU} $ and the same temperature profile \citep[][Eq. 25]{Lenz2019}. The different colors correspond to different species. Times that are not accounted for in \citet{Lenz2019} are shaded in grey. We note that \citet{Lenz2019} do not include planetesimal collisions, hence why here collisional dust and pebbles do not exist. However, they do consider radial transport and resolve particle size bins, which is not implemented in our model.}}
\label{fig:comparison_to_lenzetal2019}
\end{figure*}

\change{
In Fig. \ref{fig:comparison_to_lenzetal2019}, we compare the integrated mass of planetesimals and smaller particles in our model to \citet{Lenz2019}, where planetesimal collisions are not implemented. Moreover, \citet{Lenz2019} solve the Smoluchowski equation for growth and fragmentation of particles \citep{Smoluchowski1916}, instead of approximating dust coagulation via the two population model, which tends to underestimate growth timescales and thus overestimates particle growth \citep{Birnstiel2012}. For this reason, dust growth and therefore also planetesimal formation sets in sooner in our model compared to \citet{Lenz2019}. Further, \citet{Lenz2019} use the same prescription for planetesimal formation, which is reflected in the slope of the planetesimal mass evolution of the two models in Fig. \ref{fig:comparison_to_lenzetal2019}. However, once planetesimal collisions become relevant, i.e. when about 30 \% of the total mass available in solids is comprised in planetesimals, our model predicts a smaller planetesimal fraction than \citet{Lenz2019}.
}

\section{On treating the pebble Stokes number as a fixed parameter}
\label{sec:pebble_stokes_number}

\begin{figure*}[t]
\centering
\includegraphics[width=\linewidth]{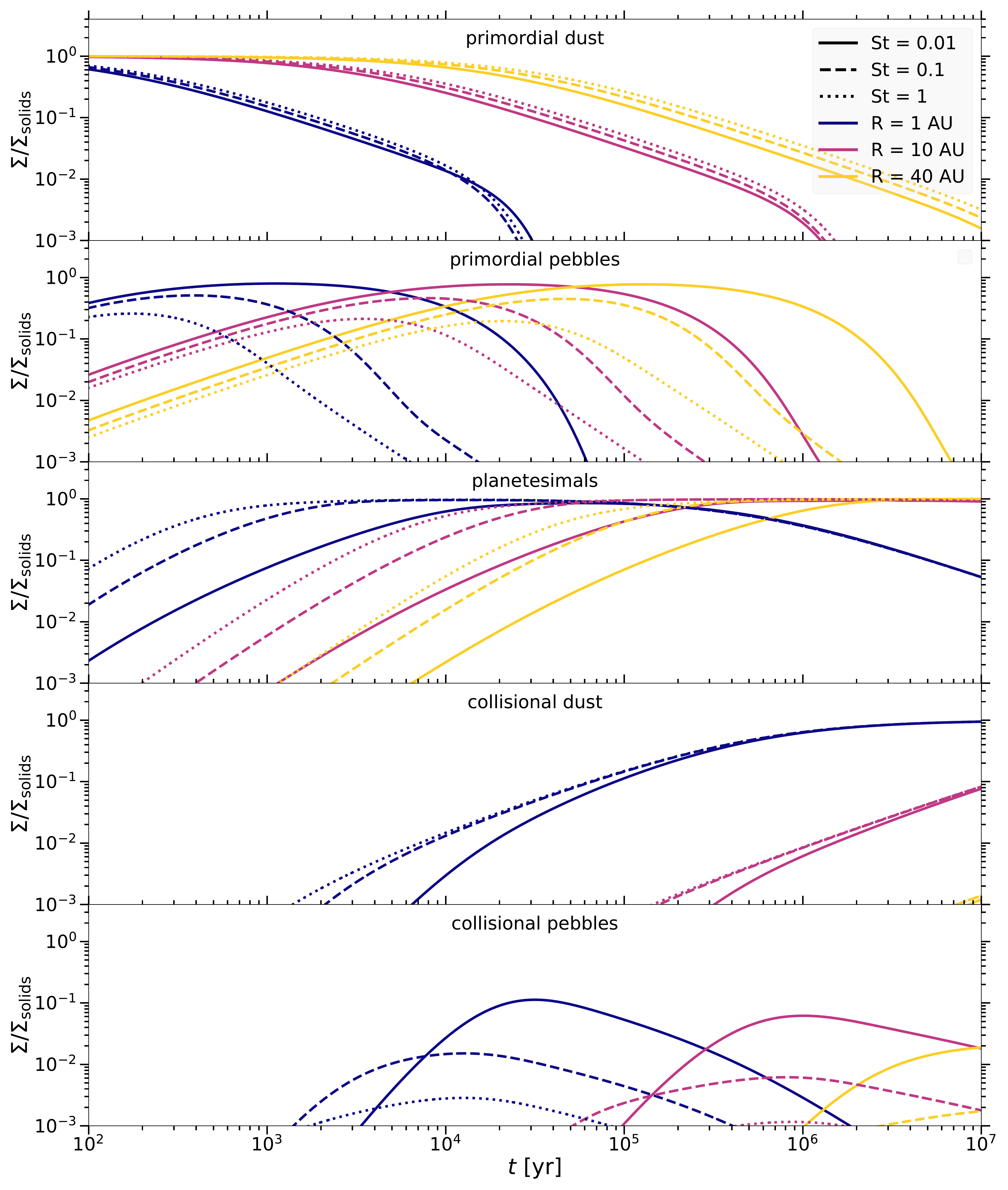}
\caption[]{\change{Local evolution at $R$ for $\epsilon = 0.01$ and $\mu = 0.01$ of normalized column density of the species for different pebble Stokes numbers (dotted lines: $\mathrm{St}_\mathrm{pbb} = 1$, solid lines (see Fig. \ref{fig:evolution_fixed_parameter}): $\mathrm{St}_\mathrm{pbb} = 0.1$, dashed lines: $\mathrm{St}_\mathrm{pbb} = 0.01$) and different radii (colors). Each Panel corresponds to a different particle species: primordial dust and pebbles, planetesimals, collisional dust and pebbles (from top to bottom).} }
\label{fig:Stokes_number_variation}
\end{figure*}

\change{Following \citet{Birnstiel2012}, the most abundant pebble size $a$ resides at the fragmentation limit (typically in the inner disk) or at the drift limit (typically in the outer disk), i.e. in terms of Stokes numbers
\begin{equation}
    \mathrm{St}_\mathrm{pbb} = \min\left(\mathrm{St}_\mathrm{frag},\mathrm{St}_\mathrm{drift}\right),
\end{equation}
with $\mathrm{St}_\mathrm{frag}$ being the maximum Stokes number in a fragmentation dominated size distribution, and $\mathrm{St}_\mathrm{drift}$ the maximum Stokes number that can be reached before drift removes the particle. In our model, we treat the pebble Stokes number as a fixed parameter with $\mathrm{St}_\mathrm{pbb} = 0.1$. This is a very good approximation for $\alpha = 10^{-3}$ \citep[see][Fig. 3]{Lenz2019}. In more turbulent disks ($\alpha \geq 10^{-2}$) the higher relative velocities of dust particles lead to a decrease of the fragmentation size and our approximation of $\mathrm{St}_\mathrm{pbb} = 0.1$ is too high, especially in the inner disk. Hence, we investigate the influence of the pebble Stokes number on the local evolution in Fig. \ref{fig:Stokes_number_variation}.}

\change{
For overestimates of the pebble Stokes number, the dust-to-pebbles growth timescale will also be overestimated, since $\tau_\mathrm{growth} \propto \ln{\left(\mathrm{St}_\mathrm{pbb}\right)}$ after Eq. \eqref{eq:adjustedgrowthtimescale}. Because of this, the pebble population in Fig. \ref{fig:Stokes_number_variation} grows faster for smaller Stokes numbers. Conversely, planetesimals form slower for smaller Stokes numbers, as we expect a decrease in drift velocity as portrayed in Eq. \eqref{eq:vdrift}. As this decrease is dominating over the increase of available pebbles, the effect of the pebble Stokes number is similar to the effect of the trapping efficiency depicted in Fig. \ref{fig:epsilon_variation}.}

\change{
In summary, for smaller pebble Stokes numbers, we expect more pebbles, though they drift significantly slower resulting in a smaller pebble flux and less planetesimals. In our parameter study, this error gains importance for disks with moderate or high turbulence, and typically is in the inner disk more significant than in the outer disk.
}

\section{Details of the planetesimal collision model}

\subsection{Normalizing the specific number column density power law}
\label{sec:collision_normalize}

\change{
In our model, colliding planetesimals are equally sized and therefore equally massive. The surface density must be conserved during a collision, i.e.
\begin{equation}
\label{eq:collision_conservation}
2 \sigma_\mathrm{pls} = \int_{m_0}^M n_m (m) m\, \mathrm{d}m = C_n \int_{m_0}^M m^{1-\xi} \mathrm{d}m,
\end{equation}
where $\sigma_\mathrm{pls}$ is a parameter specifying the surface density of a single planetesimal, and $m_0$ and $M$ the masses of the smallest and largest possible fragment respectively. We set $m_0 = 4/3 \pi a_\mathrm{0}^3$ and $M = m_\mathrm{pls}$ for destructive collisions.
We solve the integral in Eq. \eqref{eq:collision_conservation}, giving
\begin{align}
    2 \sigma_\mathrm{pls} = C_n \left[\frac{M^{2-\xi}}{2-\xi} - \frac{m_0^{2-\xi}}{2-\xi}\right]= C_n\frac{  M^{2-\xi}}{2-\xi}\left[1- \left(\frac{m_0}{M}\right)^{2-\xi}\right],
\end{align}
leading to
\begin{align}
    C_n = 2 \sigma_\mathrm{pls}\left(2-\xi\right)M^{\xi-2}\left[1- \left(\frac{m_0}{M}\right)^{2-\xi}\right]^{-1}.
\end{align}
We note that the choice of $\sigma_\mathrm{pls}$ is irrelevant, because it cancels when calculating the relative fractions $p_\mathrm{dst}, p_\mathrm{pbb}$ and $p_\mathrm{pls}$.
}

\subsection{Calculation of the transition masses $m_1$ and $m_2$}
\label{section:transition_masses}

The drift velocity of a particle such that it can just be considered a pebble and not dust, is equal to the required velocity for it to cover one trap distance \newchange{$d$} during exactly one trap lifetime
\change{\begin{equation}
    v_\mathrm{drift}(m_1) \stackrel{!}{=} \frac{d}{\tau_\mathrm{trap}}
\end{equation}}
If this \change{condition is met, a particle with mass $m_1$} has just enough time to participate in planetesimal formation via graviational instability and can therefore be considered a pebble. \change{Plugging Eq. \eqref{eq:St_epstein} into the drift velocity in Eq. \eqref{eq:vdrift} yields a second order polynomial with the smaller solution being \begin{equation}
m_1 = \frac{2 \Sigma_\mathrm{g}^3}{3\pi^2 \rho^2}\left[\frac{h_\mathrm{g}^2\gamma \tau_\mathrm{trap}\Omega}{R d}- \sqrt{\left(\frac{h_\mathrm{g}^2\gamma \tau_\mathrm{trap}\Omega}{R d}\right)^2 - 4}\right]^3,
\end{equation} where we used $(\epsilon_\mathrm{dg}+1)^2 \approx 1$.}

\change{Following \citet{Birnstiel2012}, we compare a particle's} drift timescale to its growth timescale, which yields the maximum Stokes number, that can be reached before \change{it is removed by radial drift, i.e. 
\begin{equation}
\label{eq:frag_condition_2}
\frac{R}{v_\mathrm{drift}(m_2)} \stackrel{!}{=} \tau_\mathrm{growth}.
\end{equation}
Particles more massive than $m_2$ grow faster than they are removed from their position, potentially forming planetesimals by growth and not by gravitational collapse, which is not the focus of this paper. Hence, we assign all particles with mass $m > m_2$ to the planetesimal population. The condition in \eqref{eq:frag_condition_2} also results in a quadratic equation after using Eqns. \eqref{eq:St_epstein}, \eqref{eq:adjustedgrowthtimescale} and \eqref{eq:vdrift}. We then define the larger solution as the transition of pebbles to planetesimals in the fragment distribution: 
\begin{equation}
m_2 = \frac{2 \Sigma_\mathrm{g}^3}{3\pi^2 \rho^2} \left[ \frac{\gamma h_\mathrm{g}^2}{\epsilon_\mathrm{dg}R^2} + \sqrt{\left(\frac{\gamma h_\mathrm{g}^2}{\epsilon_\mathrm{dg}R^2}\right)^2 - 4}\right]^3.
\end{equation}
}

\end{document}